\documentclass[nofootinbib,aps,11pt]{revtex4}
\usepackage{graphicx}
\usepackage{cancel}
\usepackage{amssymb}
\usepackage{textcomp}
\usepackage{amsmath}
\usepackage{bm}
\usepackage{times}
\usepackage{epsfig}
\usepackage{color}
\usepackage{axodraw}

\begin{document}

\title{\Large 
Non-diagonal Charged 
Lepton Yukawa Matrix: \\
Effects on Neutrino Mixing in Supersymmetry }
\author{Giovanna Cottin}
\author{Marco Aurelio D\'\i az}
\author{Benjamin Koch}
\affiliation{
{\small Departamento de F\'\i sica, Pontificia Universidad Cat\'olica 
de Chile, Avenida Vicu\~na Mackenna 4860, Santiago, Chile 
}}
\begin{abstract}
Generally the diagonalization of the mass matrix of the charged leptons
is a part of the neutrino $U_{PMNS}$ matrix.
However, usually this contribution is ignored by assuming
a diagonal mass matrix for charged leptons.
In this letter we test this common assumption in
the context of neutrino physics.
Our analytical and numerical results for 
two supersymmetric models
reveal that such a simplification is not justified.
Especially for the solar and reactor mixing angles important
modifications are found.
\end{abstract}
\date{\today}
\maketitle
\section{Introduction}
\label{sec_Intro}

Supersymmetric models which incorporate small violations
of R-parity \cite{Barbier:2004ez} 
are of special interest in the context of neutrino
phenomenology \cite{Hempfling:1995wj}. It has been shown that they can give rise
to neutrino masses and mixing angles that are compatible
with experimental data. In specific models this
is achieved by either taking into account
low scale gravity effects, or by including
loop effects in the neutrino propagator
\cite{hep-ph/0302021,Diaz:2009yz,Diaz:2009gf,arXiv:1106.0308,arXiv:1109.0512}.

While neutrino masses and mixings are considered
``new physics'' beyond the Standard Model (SM),
the masses, mixing angles and phase of the other nine
fermions are described, within the SM,
by using thirteen independent parameters.
It has however been pointed out that in models
motivated by supersymmetric gauge unification, 
the number of free parameters can be reduced to eight
\cite{Georgi:1979df,Dimopoulos:1991za}.
In those Grand Unified Theories (GUT), also the lepton mass matrix 
is non-diagonal and therefore has to be diagonalized
in order to reproduce the observed charged lepton 
masses $m_e, m_\mu, m_\tau$.

\begin{figure}[h]
\centerline{\protect\vbox{\epsfig{file=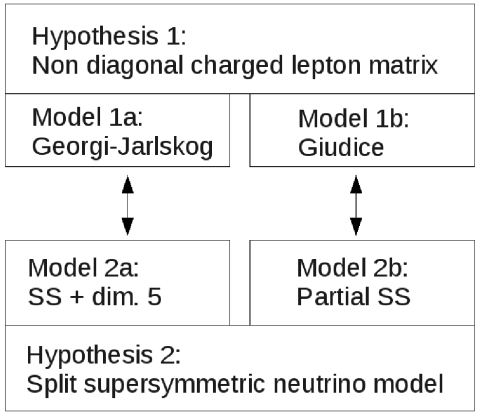,width=0.4\textwidth}}}
\caption{\it Conceptual flow chart of how the basic ideas of
a non-diagonal charge lepton Yukawa matrix and 
split-supersymmetric neutrino models are combined and studied.
}
\label{f0}
\end{figure}

Given the success of those approaches,
it is natural to seek the combination of the supersymmetric description
for the neutrino sector with the supersymmetric description
of the mass sector of the other fermions.
Since the well known neutrino $U_{PMNS}$ matrix contains also a part
that originates from the charged fermion mass sector 
(studied only in few cases \cite{Krolikowski:2005rj},
and neglected in most cases), a combination
of the neutral and charged fermion sectors is
typically not trivial. 
In this paper it will be studied how at low energy
the GUT fermion mass matrices \cite{Georgi:1979df,Dimopoulos:1991za}
affect the predictions of  neutrino 
models with R-parity violation \cite{Diaz:2009yz,Diaz:2009gf}.
The models studied in the neutrino context are Split Supersymmetry (SS)
and Partial Split Supersymmetry (PSS), linked each to corresponding
well known examples of charged lepton mass matrix textures.
The conceptual frame of this combination of
two viable and successful ideas and its realization
in terms of explicit models is shown in figure \ref{f0}.
%
A priory there is no restriction
when combining models for charged leptons
with models for neutrinos.
We decided to study two specific examples as shown in figure$\,$\ref{f0}.

\section{NEUTRAL AND CHARGED FERMIONS IN BRPV}
\label{sec_BRpV}

In the supersymmetric models we are studying here, tree-level contributions 
to neutrino masses and mixings
arise from the neutrino-neutralino mixing due to bilinear R-parity violation.
In general, when writing down the gauge invariant terms that violate R-parity
one can consider Lagrange terms that contain three fields (trilinear)
and terms that contain two fields (bilinear).
In the context of SS all the trilinear terms are irrelevant since
they contain heavy scalars that are integrated out of the effective
theory.

In BRpV models neutralinos mix with neutrinos such that a $7\times7$ mass matrix
is generated. In the base 
$\psi_0^T=(-i\widetilde B,i\widetilde W^0,\widetilde H_d^0, \widetilde H_u^0,
\nu_e, \nu_\mu, \nu_\tau)$
the corresponding terms in the lagrangian are grouped as,
\begin{equation}
{\cal L}_N = -\frac{1}{2}(\psi^0)^T {\cal M}_N \psi^0
\end{equation}
with the mass matrix introduced in blocks \cite{hep-ph/0302021},
\begin{equation}
{\cal M}_N=\left(\begin{array}{cc} {\mathrm M}_{\chi^0} & m^T \\ 
m & 0 \end{array}\right),
\label{X07x7}
\end{equation}
This neutralino/neutrino mass matrix is diagonalized with the rotation matrix,
\begin{equation}
{\cal N} = \left( \begin{matrix} N & 0 \cr 0 & U_\nu^T \end{matrix} \right)
\left( \begin{matrix} 
1-\frac{1}{2}\xi^T \xi & \xi^T \cr -\xi & 1-\frac{1}{2}\xi \xi^T \end{matrix} \right)
\equiv
\left( \begin{matrix} N & 0 \cr 0 & U_\nu^T \end{matrix} \right) {\cal N}_\xi
\end{equation}
with $\xi=m {\mathrm M}_{\chi^0}^{-1}$ at first order in perturbation theory. 
The matrix ${\cal N}_\xi$ allows a block diagonalization such that,
\begin{equation}
{\cal N}_\xi {\cal M}_N {\cal N}_\xi^T =
\left(\begin{matrix} {\mathrm M}_{\chi^0} & 0 \cr 0 & {\mathrm M}_{\nu}^{eff}
\end{matrix} \right)
\label{BlockD}
\end{equation}
with ${\mathrm M}_{\nu}^{eff}=-m {\mathrm M}_{\chi^0}^{-1} m^T$.
Matrices $N$ and $U_\nu$ further diagonalize the neutralino mass matrix 
${\mathrm M}_{\chi^0}$ and the effective neutrino mass matrix 
${\mathrm M}_{\nu}^{eff}$ respectively:
\begin{equation}
\left(
\begin{array}{cc}
 N&0\\
0&U^T_\nu
\end{array}
\right)
\left(
\begin{array}{cc}
 {\mathcal{M}}_{\bar \chi^0}&0\\
0&{\mathcal{M}}_{\nu}^{eff}
\end{array}
\right)
\left(
\begin{array}{cc}
 N^T&0\\
0&U_\nu
\end{array}
\right)=
\left(
\begin{array}{cc}
 {\mathcal{M}}_{\bar \chi^0}^{diag}&0\\
0&{\mathcal{M}}_{\nu}^{diag}
\end{array}
\right)
\end{equation}
We call these eigenstates
$F^0_i$ with $i=1,...7$.

As we will see, in order to correctly define the neutrino mixing angles 
we need to study the charged lepton sector as well. In BRpV charginos mix with 
charged leptons forming the following mass terms,
\begin{equation}
{\cal L}_C = -\frac{1}{2} \left( \psi^{+T} , \psi^{-T} \right)
\left( \begin{matrix} 0 & {\cal M}_C^T \cr {\cal M}_C & 0 \end{matrix} \right)
\left( \begin{matrix} \psi^+ \cr \psi^- \end{matrix} \right)
\end{equation}
where the basis is 
$\psi^{-T}=(-i\widetilde W^-, \widetilde H_d^-, e_L^-, \mu_L^-, \tau_L^-)$ and
$\psi^{-T}=(-i\widetilde W^+, \widetilde H_u^+, e_R^+, \mu_R^+, \tau_R^+)$. We
divide the $5\times5$ mass matrix into blocks \cite{hep-ph/0302021},
\begin{equation}
{\cal M}_C = \left( \begin{matrix}
M_{\chi^+} & Y \cr m_c & M_\ell
\end{matrix} \right)
\label{BlockC}
\end{equation}
This chargino/charged lepton mass matrix is not symmetric thus it is diagonalized by 
two matrices
\begin{equation}
{\cal U} {\cal M}_C {\cal V}^T = {\cal M}_C^{diag}
\end{equation}
where we first look for a block diagonalization, as in the neutral case, performed
by matrices ${\cal U}_\xi$ and ${\cal V}_\xi$. Neglecting $Y$ (small Yukawa 
couplings and sneutrino vevs) we find,
\begin{equation}
{\cal U}_\xi = \left( \begin{matrix}
1-\frac{1}{2} \xi_L^T \xi_L & \xi_L^T \cr
-\xi_L & 1-\frac{1}{2} \xi_L \xi_L^T
\end{matrix} \right)
\,,\qquad
{\cal V}_\xi = \left( \begin{matrix}
1-\frac{1}{2} \xi_R^T \xi_R & \xi_R^T \cr
-\xi_R & 1-\frac{1}{2} \xi_R \xi_R^T
\end{matrix} \right)
\end{equation}
with $\xi_L=m_c M_{\chi^+}^{-1}$ and 
$\xi_R=M_{\ell} m_c M_{\chi^+}^{-1} (M_{\chi^+}^{-1})^T$. In the small lepton masses 
and small BRpV parameters approximation, $\xi_R$ can
be neglected. This implies that to first order on BRpV parameters the chargino and 
the charged lepton mass matrices are unchanged by the block diagonalization,
\begin{equation}
M_{\chi^+}^{eff} = M_{\chi^+} \,,\qquad M_{\ell}^{eff} = M_{\ell}
\end{equation}
The full diagonalization is accomplished with,
\begin{equation}
{\cal U} = \left( \begin{matrix} U & 0 \cr 0 & V_L \end{matrix} \right) {\cal U}_\xi
\,,\qquad
{\cal V} = \left( \begin{matrix} V & 0 \cr 0 & V_R \end{matrix} \right) {\cal V}_\xi
\end{equation}
where
\begin{equation}
U M_{\chi^+} V^T = M_{\chi^+}^{diag} \,,\qquad
V_L M_{\ell} V_R^T = M_{\ell}^{diag}
\end{equation}
The matrices $M_{\chi^+}^{diag}$ and $M_{\ell}^{diag}$ contain the final chargino 
and charged lepton masses. We call these eigenstates $F^\pm_i$ with $i=1,...5$.

\section{GUT motivated Ansatz for Charged Leptons Mass Matrix}
\label{sec_GUT}

Grand Unified Theories provide a well motivated framework to study non-diagonal
charged lepton mass matrices. The most studied Grand Unification gauge groups 
are $SU(5)$ and $SO(10)$, which break down to the SM gauge group 
$SU(3)\times SU(2)\times U(1)$. In addition, the GUT can be embedded into 
supersymmetry. In this context, different proposals for a charged
lepton mass matrix are postulated at the GUT scale. In the following subsections 
we will study two GUT examples based on the two groups mentined above.

\subsection{Georgi-Jarlskog Ansatz}
\label{sub_GJ}

We consider first the Georgi-Jarlskog ansatz \cite{Georgi:1979df} for the
charged lepton mass matrix, introduced in the context of an $SU(5)$ GUT theory, 
and re-analyzed in \cite{Dimopoulos:1991za} for a supersymmetric $SO(10)$ GUT
group. Written in the notation of the later article, the charged lepton mass 
matrix depends on three parameters $D$, $E$, and $F$, which we assume real. We 
have,
\begin{equation}
M_\ell = \frac{v}{\sqrt{2}} \left(
\begin{matrix} 0 & F & 0 \cr F & -3E & 0 \cr 0 & 0 & D \end{matrix}
\right)
\label{GJansatz}
\end{equation}
which essentially does not change after RGE running effects. 
The matrix is proportional to $v$ when the low energy theory contains only one 
Higgs doublet (for example Split Supersymmetry). In the case it contains two 
Higgs doublets (for example Partial Split Supersymmetry) the replacement 
$v\rightarrow v_d$ must be made.

If we assume $E$ is positive, the eigenvalues are,
\begin{eqnarray}
m_{\ell_1} &=& \frac{v}{2\sqrt{2}} \left( -3E+\sqrt{9E^2+4F^2} \right)
\nonumber\\
m_{\ell_2} &=& \frac{v}{2\sqrt{2}} \left( -3E-\sqrt{9E^2+4F^2} \right)
\\
m_{\ell_3} &=& \frac{v}{\sqrt{2}} D
\nonumber
\end{eqnarray}
These eigenvalues, up to a possible sign, are equal to $m_e=0.511$ MeV,
$m_\mu=105.7$ MeV, and $m_\tau=1777$ MeV respectively \cite{Nakamura:2010zzi},
fixing the parameters in the charged lepton Yukawa matrix to 
$F=4.22\times 10^{-5}$, $E=2.01\times 10^{-4}$, and $D=1.02\times 10^{-2}$. 

 It is clear that only one angle is enough to parametrize 
$M_{\ell}^{diag}=V_L M_{\ell} V_R^T$. Since $M_{\ell}$ is symmetric, the 
diagonalization matrix has the following form,
\begin{equation}
V_L = V_R = \left(
\begin{matrix}  \cos\alpha & \sin\alpha & 0 \cr 
               -\sin\alpha & \cos\alpha & 0 \cr 
               0               &  0               & 1 \end{matrix}
\right)\,,\qquad
\tan2\alpha = \frac{2F}{3E}
\label{GJ}
\end{equation}
This angle is such that $\sin\alpha\approx0.0695$.

\subsection{Giudice Ansatz}
\label{sub_G}

The second ansatz we consider was introduced by G.~Giudice \cite{Giudice:1992an}
in the context of supersymmetric Grand Unified Theories (GUT). The charged
lepton mass matrix is,
\begin{equation}
\label{ansatz2}
M'_\ell = \frac{v_d}{\sqrt{2}} \left(
\begin{matrix} 0 & F & 0 \cr F & -3E & 2E \cr 0 & 2E & D \end{matrix}
\right)
\end{equation}
whose Yukawa couplings do not change after running.
The implications of this type of ansatz in terms of neutrino
textures have been investigated in \cite{hep-ph/9409369,hep-ph/9509351,hep-ph/0012046}.
We will associate this ansatz
with Partial Split Supersymmetry, hence the mass matrix is proportional to $v_d$.
The hierarchical nature of the charged lepton and quarks necessitates $F\ll E\ll D$. 
In this approximation we find the following eigenvalues,
\begin{eqnarray}
m_{\ell_1} &\approx& \frac{v_d}{\sqrt{2}} \, \frac{F^2}{3E}
\nonumber\\
m_{\ell_2} &\approx& \frac{v_d}{\sqrt{2}} \left( -3E -4\frac{E^2}{D} \right)
\\
m_{\ell_3} &\approx& \frac{v_d}{\sqrt{2}} \left( D +4\frac{E^2}{D} \right)
\nonumber
\end{eqnarray}
Imposing the experimental values of the charged leptons into these results we
find $Fc_\beta=4.17\times 10^{-5}$, $Ec_\beta=1.97\times 10^{-4}$, and 
$Dc_\beta=1.02\times 10^{-2}$. Note that these Yukawa parameters grow with 
$\tan\beta$. Notice also that the numerical value of the parameters $F$, $E$,
and $C$ differ only slightly with respect to the ones obtained for the 
previous ansatz (for $v=v_d \Leftrightarrow \cos \beta=1$). This is related to
the fact that the
charged 
lepton masses are hierarchical. 

The mass matrix $M'_\ell$ in eq.~(\ref{ansatz2}) is diagonalized by the following
matrix,
\begin{equation}
V'_L = V'_R \approx \left(
\begin{matrix}  1            & -\frac{F}{3E} & 0 \cr 
                \frac{F}{3E} &  1            & \frac{2E}{D}    \cr 
                0            & -\frac{2E}{D} & 1               \end{matrix}
\right)
\end{equation}
where we have neglected smaller terms. We parametrize this rotation matrix with
two angles,
\begin{equation}
V_L' = \left(
\begin{matrix}  \cos\alpha' & \sin\alpha' & 0 \cr 
               -\sin\alpha' & \cos\alpha' & 0 \cr 
               0              & 0             & 1 \end{matrix}
\right)
\left(
\begin{matrix} 1 &  0             & 0             \cr 
               0 &  \cos\theta' & \sin\theta' \cr 
               0 & -\sin\theta' & \cos\theta' \end{matrix}
\right)
\,,\quad \tan2\alpha' \approx \frac{2F}{3E}
\,,\quad \tan2\theta' \approx -\frac{4E}{D}
\label{Giudice}
\end{equation}
These angles are such that $\sin\alpha'\approx0.070$ and $\sin\theta'\approx0.036$.
Notice that $V_L'(\theta'=0, \alpha'=\alpha)=V_L(\alpha)$.

\section{${\bf U_{PMNS}}$ and ${\bf W}$ Boson Coupling to Fermions}
\label{sec_ncW}

Charged and neutral fermion couplings to the $W$ boson are essential
for the $U_{PMNS}$ matrix of neutrino mixing angles
because they define the base where charged leptons are diagonal. 
In BRpV models
the situation
is complicated by the fact that charginos mix with charged leptons, as we saw 
in the previous chapter. The relevant coupling is,
\begin{center}
\vspace{-50pt} \hfill \\
\begin{picture}(110,90)(0,23) 
\Photon(10,25)(50,25){3}{4}
\ArrowLine(50,25)(78,53)
\ArrowLine(78,-3)(50,25)
\Text(10,35)[]{$W^-$}
\Text(90,55)[]{$F_i^-$}
\Text(90,-5)[]{$F_j^0$}
\end{picture}
$
=\,i\,\gamma^\mu\,\Big[O^{cnw}_{Lij}\frac{(1-\gamma_5)}{2}+
O^{cnw}_{Rij}\frac{(1+\gamma_5)}{2}\Big]
$
\vspace{30pt} \hfill \\
\end{center}
\vspace{10pt}
with
\begin{eqnarray}
O_{Lij}^{cnw} &=& -g \bigg[ {\cal N}_{j2} \,{\cal U}_{i1} + \frac{1}{\sqrt{2}}
\Big( {\cal N}_{j3} \,{\cal U}_{i2} + 
\sum_{k=1}^3 {\cal N}_{j,4+k} \,{\cal U}_{i,2+k} \Big) \bigg]
\nonumber\\
O_{Rij}^{cnw} &=& -g \bigg[ {\cal N}_{j2} \,{\cal V}_{i1} + \frac{1}{\sqrt{2}}\,
{\cal N}_{j4} \,{\cal V}_{i2} \bigg]
\end{eqnarray}
In first approximation in $\epsilon/M_\chi$ we use,
\begin{equation}
{\cal N} = \left( \begin{matrix} N & N \xi^T \cr -U_\nu^T \xi & U_\nu^T 
\end{matrix} \right)
\,,\qquad
{\cal U} = \left( \begin{matrix} U & U \xi_L^T \cr -V_L \xi_L & V_L 
\end{matrix} \right)
\,,\qquad
{\cal V} = \left( \begin{matrix} V & 0 \cr 0 & V_R \end{matrix} \right)
\end{equation}
and find for the charged lepton and neutrino coupling to $W$ Bosons the
following,
\begin{center}
\vspace{-50pt} \hfill \\
\begin{picture}(110,90)(0,23) 
\Photon(10,25)(50,25){3}{4}
\ArrowLine(50,25)(78,53)
\ArrowLine(78,-3)(50,25)
\Text(10,35)[]{$W^-$}
\Text(90,55)[]{$\ell_i^-$}
\Text(90,-5)[]{$\overline\nu_j$}
\end{picture}
$
=\,-i\,\frac{g}{\sqrt{2}}\,\left(V_LU_\nu\right)_{ij}\,
\gamma^\mu\,\frac{(1-\gamma_5)}{2}
$
\vspace{30pt} \hfill \\
\end{center}
\vspace{10pt}
Therefore, the neutrino mixing angles are defined by,
\begin{equation}
U_{PMNS} = V_L U_\nu
\end{equation}
Notice that the $U_{PMNS}$ matrix coincides with 
the matrix that diagonalizes the neutrino mass matrix, $U_\nu$, only 
when the charged leptons are diagonal in the original basis. Otherwise, there
is an extra contribution from the left matrix $V_L$ that diagonalizes the charged 
lepton mass matrix.

Using the following convention for the neutrino angles,
\begin{equation}
U_{PMNS}=
\left(\begin{matrix} 1 &  0      & 0      \cr 
                     0 &  c_{23} & s_{23} \cr 
                     0 & -s_{23} & c_{23} \end{matrix}\right)
\left(\begin{matrix}  c_{13}             & 0 & s_{13}e^{i\delta} \cr 
                      0                  & 1 & 0                 \cr 
                     -s_{13}e^{-i\delta} & 0 & c_{13}            \end{matrix}\right)
\left(\begin{matrix}  c_{12} & s_{12} & 0 \cr 
                     -s_{12} & c_{12} & 0 \cr 
                      0      & 0      & 1 \end{matrix}\right)
\label{U}
\end{equation}
and assuming this matrix is real ($\delta=0$), the general structure for 
the mixing angles considering the  Giudice ansatz for the charged leptons
(\ref{Giudice}) is given by,
\begin{eqnarray}
\sin\theta_{13}^{(V'_L\ne1)} &=& \sin\theta_{13}^{(V'_L=1)} + s'_\alpha
s_{23} c_{13}
\nonumber\\
\tan\theta_{23}^{(V'_L\ne1)} &=& \tan\theta_{23}^{(V'_L=1)} \left\{
1+s'_\theta \left( t_{23}+\frac{1}{t_{23}} \right) - s'_\alpha \frac{t_{13}}{s_{23}}
\right\}
\nonumber\\
\tan\theta_{12}^{(V'_L\ne1)} &=& \tan\theta_{12}^{(V'_L=1)} \left\{
1+s'_\alpha \frac{c_{23}}{c_{13}} \left( t_{12}+\frac{1}{t_{12}} \right)
\right\}
\label{anlgeschange}
\end{eqnarray}
where we have used the fact that the angles $\alpha'$ and $\theta'$ are small.
Analogous expressions for the Georgi-Jarlskog ansatz are obtained by the
substitution $V'_L\rightarrow V_L$, $\theta'=0$, and $\alpha'\rightarrow\alpha$.

\section{Split Supersymmetry with Flavor Blind Dimension Five}
\label{sec_SS}

In Split Supersymmetry \cite{ArkaniHamed:2004fb} all scalars are very heavy, for 
simplicity degenerated at a mass $\tilde{m}$, except for one Higgs doublet. 
Integrating out the heavy scalars the SS Lagrangian includes,
\begin{align}
{\mathcal{L}}_{SS}&\owns 
-\Big[ m^{2}H^{\dag}H + \frac{\lambda}{2}(H^{\dag}H)^{2}\Big] - 
Y_{u}\overline{Q}_{L}u_{R}i\sigma_{2}H^{*} - Y_{d}\overline{Q}_{L}d_{R}H 
-Y_{e}\overline{L}_{L}e_{R}H\nonumber\\[0.1cm]& - \frac{M_{3}}{2}\tilde{G}\tilde{G}- 
\frac{M_{2}}{2}\tilde{W}\tilde{W} - \frac{M_{1}}{2}\widetilde{B}\widetilde{B}- 
\mu\widetilde{H}^{T}_{u}i\sigma_{2}\widetilde{H}_{d}- 
\frac{1}{\sqrt{2}}H^{\dag}(\widetilde{g}_{u}\sigma\widetilde{W} + 
\tilde{g}'_{u}\widetilde{B})\widetilde{H}_{u}\nonumber\\[0.1cm]&- 
\frac{1}{\sqrt{2}}H^{T}i\sigma_{2}(-\widetilde{g}_{d}\sigma\widetilde{W} + 
\widetilde{g}'_{d}\widetilde{B})\widetilde{H}_{d} + \mathrm{h.c},
\end{align}
The last two terms are the Higgs-gaugino-higgsino interactions, with couplings 
$\tilde g$ induced by integrating out the heavy scalars.

Split Supersymmetry with violation of R-Parity \cite{Diaz:2006ee} includes the 
extra terms
\begin{equation}
{\cal L}_{SS}^{RpV} \owns
\epsilon_i\widetilde H^T_u i\sigma_2 L_i
-\frac{1}{\sqrt{2}} a_i H^T i \sigma_2
(-\tilde g_d \sigma\widetilde W+\tilde g'_d\widetilde B)L_i \ + \ \mathrm{h.c.}.
\label{LSplitRpV}
\end{equation}
The first term corresponds to the usual bilinear violation of R-Parity,
which mixes higgsinos with leptons through the mass parameters $\epsilon_i$.
The terms proportional to the $a_i$ parameters are generated as effective terms 
in the SS lagrangian after integrating out the sfermions.

\subsection{Neutrinos and Neutralinos in SS}
\label{subsec_neutral_SS}

Now we specify the neutrino-neutralino mixing described in section \ref{sec_BRpV} 
for the Split Supersymmetric case. The upper left block  in eq.~(\ref{BlockD})  
corresponds to the neutralino sector,
\begin{equation}
{\bf M}_{\chi^0}^{SS}=\left(\begin{array}{cccc}
M_1 & 0 & -\frac{1}{2}\tilde g'_d v & \frac{1}{2}\tilde g'_u v \\
0 & M_2 & \frac{1}{2}\tilde g_d v & -\frac{1}{2}\tilde g_u v \\
-\frac{1}{2}\tilde g'_d v & \frac{1}{2}\tilde g_d v & 0 & -\mu \\
\frac{1}{2}\tilde g'_u v & -\frac{1}{2}\tilde g_u v & -\mu & 0
\end{array}\right)
\label{X0massmat}
\end{equation}
where $M_1, M_2$ are the gaugino masses, $\mu$ is the higgsino mass, and $v=246$ GeV 
is the Higgs vacuum expectation value. The neutralino/neutrino mixing is equal to,
\begin{equation}
m^{SS}=\left(\begin{array}{cccc}
-\frac{1}{2} \tilde g'_d a_1v & \frac{1}{2} \tilde g_d a_1v 
& 0 &\epsilon_1 \cr
-\frac{1}{2} \tilde g'_d a_2v & \frac{1}{2} \tilde g_d a_2v&0 
& \epsilon_2 \cr
-\frac{1}{2} \tilde g'_d a_3v & \frac{1}{2} \tilde g_d a_3v&0 
& \epsilon_3
\end{array}\right).
\end{equation}
with $\epsilon_i$ and $a_i$ the BRpV parameters described in eq.~(\ref{LSplitRpV}).
Therefore, in Split Supersymmetry the effective neutrino mass matrix is given by,
\begin{equation}
{\bf M}_\nu^{eff}=-m^{SS}\,({\mathrm{M}}_{\chi^0}^{SS})^{-1}\,(m^{SS})^T=
\frac{v^2}{4\det{M_{\chi^0}^{SS}}}
\left(M_1 \tilde g^2_d + M_2 \tilde g'^2_d \right)
\left(\begin{array}{cccc}
\lambda_1^2        & \lambda_1\lambda_2 & \lambda_1\lambda_3 \cr
\lambda_2\lambda_1 & \lambda_2^2        & \lambda_2\lambda_3 \cr
\lambda_3\lambda_1 & \lambda_3\lambda_2 & \lambda_3^2
\end{array}\right),
\label{treenumass}
\end{equation}
with $\lambda_i= a_i\mu+\epsilon_i$. The determinant of the neutralino mass 
matrix is found to be,
\begin{equation}
\det{M_{\chi^0}^{SS}}=-\mu^2 M_1 M_2 + \frac{1}{2} v^2\mu \left( 
M_1 \tilde g_u \tilde g_d + M_2 \tilde g'_u \tilde g'_d \right)
+\textstyle{\frac{1}{16}} v^4 
\left(\tilde g'_u \tilde g_d - \tilde g_u \tilde g'_d \right)^2.
\label{detNeut}
\end{equation}
For our numerical calculations we neglect the running of the $\tilde g$
couplings.

Since the effective neutrino mass matrix has only one non-zero eigenvalue, 
at tree level only the atmospheric mass squared is generated, and the solar
mass squared difference remains null. In Split Supersymmetry this does not
change when we add quatum corrections to the neutrino mass matrix. This
is a well known fact in BRpV Split Supersymmetry \cite{Davidson:2000ne}. 
Nevertheless, it has been noticed that gravity contributions via dimension 5 
operators, can generate a solar mass when the operator is suppressed by a 
reduced Planck mass, as in models with extra dimensions \cite{Berezinsky:2004zb}. 
Following ref.~\cite{Diaz:2009yz}, we include a contribution to the neutrino 
mass matrix induced by gravity,
\begin{equation}
M_\nu^G = \mu_g \left(
\begin{matrix} 1 & 1 & 1 \cr 1 & 1 & 1 \cr 1 & 1 & 1 \end{matrix}
\right)
\label{gravContr}
\end{equation}
where $\mu_g\sim v^2/M_X$ parametrizes the size of the contribution. This
parameter has units of mass, is proportional to the Higgs vacuum expectation 
value squared $v^2$, and inversely proportional to the reduced Planck mass 
$M_X$. The equality of all entries in the matrix symbolizes the expected flavor
blindness of the gravitational interactions.

Assuming that the charged lepton mass matrix is already diagonal, it was shown
in ref. \cite{Diaz:2009yz} that neutrino mass squared differences 
predict values $\mu_g\sim 3\times 10^{-3}$ eV. This corresponds to a reduced
Planck mass $M_X\sim 2\times 10^{16}$ GeV, remarkably close to the GUT mass scale. 
In addition, maximal atmospheric mixing predicts $\sin^2\theta_{sol}=1/3$, well
within the $3\sigma$ experimental result $\sin^2\theta_{sol}=0.305\pm0.075$. In 
the following we will explore the effects of a non-diagonal charged lepton mass
matrix.

\subsection{Charged Leptons and Charginos in SS}
\label{subsec_charged_SS}

In SS the chargino block in equation (\ref{BlockC}) has the following structure,
\begin{equation}
M_{\chi^+} = \left( \begin{matrix}
M_2 & \frac{1}{\sqrt{2}}\tilde g_u v \cr \frac{1}{\sqrt{2}}\tilde g_d v & \mu
\end{matrix} \right)
\label{chaSS}
\end{equation}
with all the parameters already defined in the previous sections. The charged 
lepton mass matrix has the usual form $M_{\ell}^{ij}=Y_\ell^{ij} v / \sqrt{2}$, 
with $Y_\ell$ the lepton Yukawa couplings. 

The mixing 
between charginos and charged leptons is given by the matrices
\begin{equation}
m_c = \left( \begin{matrix}
\frac{1}{\sqrt{2}}\tilde g_d a_1 v & -\epsilon_1 \cr
\frac{1}{\sqrt{2}}\tilde g_d a_2 v & -\epsilon_2 \cr
\frac{1}{\sqrt{2}}\tilde g_d a_3 v & -\epsilon_3 \cr
\end{matrix} \right)
\end{equation}
and
\begin{equation}
Y = \left( \begin{matrix}
0 & 0 & 0 \cr
-\frac{1}{\sqrt{2}}Y_\ell^{1i} a_i v & -\frac{1}{\sqrt{2}}Y_\ell^{2i} a_i v &
-\frac{1}{\sqrt{2}}Y_\ell^{3i} a_i v  \cr
\end{matrix} \right)
\end{equation}
Since lepton masses are so much smaller than chargino masses and R-Parity violating
terms $a_i$ are also small, it is usually a good approximation to neglect the 
effect of the matrix $Y$, as we did in the diagonalization process in section \ref{sec_BRpV}.

Notice that the charged lepton Yukawa matrix does not need to be diagonal.
This point is not trivial, and has consequences on the neutrino mixing angles 
as we will see in the next chapters. 

\subsection{Effects on Neutrino Parameters in SS}
\label{subsec_parameters_SS}

The effective neutrino mass matrix, including BRpV terms and a gravity induced 
contribution from a flavor blind dimension 5 operator in models with extra 
dimensions, is
\begin{equation}
\label{eq_mixmod}
M_\nu^{ij}=A_\lambda \lambda^i\lambda^j+\mu_g \,.
\end{equation}
It is obtained by summing eqs.~(\ref{treenumass}) and (\ref{gravContr}), with
the coefficient $A_\lambda$ being read from eq.~(\ref{treenumass}). The neutrino mass
parameters are not changed by the charged lepton contributions  \cite{Diaz:2009yz},
\begin{eqnarray}
\Delta m^2_{sol} &=& \mu_g^2
\frac{|\vec{v}\times\vec{\lambda}|^4}{|\vec{\lambda}|^4}+{\mathcal{O}}(\mu_g^3)
\nonumber\\
\Delta m^2_{atm} &=& A_\lambda^2|\vec\lambda|^4 + 
2A_\lambda\mu_g (\vec v\cdot\vec\lambda)^2
+ {\mathcal{O}}(\mu_g^2)
\label{mases}
\end{eqnarray}
but the mixing angles are corrected. Considering the diagonalization matrix
from Georgi-Jarlskog ansatz in eq.~(\ref{GJ}), and using the convention in 
eq.~(\ref{U}) for neutrino angles, we find
\begin{eqnarray}
\sin\theta_{13} &=& \frac{c\lambda_1+s\lambda_2}{|\vec\lambda|}
\label{anglesSS}\\
\tan\theta_{23} &=& \frac{c\lambda_2-s\lambda_1}{\lambda_3}
\nonumber\\
\tan\theta_{12} &=& \frac{1}{|\vec\lambda|} \left[
\frac{c(\lambda_2^2+\lambda_3^2-\lambda_1\lambda_2-\lambda_1\lambda_3)
     +s(\lambda_1^2+\lambda_3^2-\lambda_1\lambda_2-\lambda_2\lambda_3)}
{c(\lambda_3-\lambda_2)+s(\lambda_1-\lambda_2)} \right]
\nonumber
\end{eqnarray}
where $c=\cos\alpha$ and $s=\sin\alpha$. These relations are a special case
of the general formulae in eq.~(\ref{anlgeschange}).

From this we can learn the following. First, the $3\sigma$ upper bound 
$\sin^2\theta_{13}<0.035$ \cite{Schwetz:2011zk} implies that a good approximation, as in 
\cite{Diaz:2009yz}, is $\lambda_1^2\ll\lambda_2^2+\lambda_3^2$. Therefore, the 
correction on $\sin^2\theta_{13}$ may be very significant since both terms in
$c\lambda_1+s\lambda_2$ are comparable. Second, the correction on the 
atmospheric angle is of a second order, since $s$ and $\lambda_1/|\vec\lambda|$
are small. Third, the correction on the solar angle is typically of the order of
$s$ ($\sim7\%$), which is not negligible. 
In section \ref{sec_Numerical} these effects
are studied numerically.

\section{Partial Split Supersymmetry}
\label{sec_PSS}

In Partial Split Supersymmetry all sfermions are heavy, for simplicity degenerate 
with a mass $\widetilde m$, while the two Higgs doublets remain at the weak
scale \cite{Diaz:2006ee,Diaz:2009gf,Sundrum:2009gv}. The lagrangian includes,
\begin{eqnarray}
{\cal L}_{PSS} &\owns& - \Big[ m_1^2H_d^\dagger H_d 
+ m_2^2H_u^\dagger H_u - m_{12}^2(H_d^T\epsilon H_u+h.c.)
\nonumber\\ &&
+ \textstyle{\frac{1}{2}}\lambda_1(H_d^\dagger H_d)^2 
+ \textstyle{\frac{1}{2}}\lambda_2(H_u^\dagger H_u)^2 
+ \lambda_3(H_d^\dagger H_d)(H_u^\dagger H_u)
+ \lambda_4|H_d^T\epsilon H_u|^2
\Big]
\nonumber\\ &&
+ Y_u \overline u_R H_u^T \epsilon q_L 
- Y_d \overline d_R H_d^T \epsilon q_L
- Y_e \overline e_R H_d^T \epsilon l_L
\label{LagSS2HDM}\\ &&
-\textstyle{\frac{1}{\sqrt{2}}} H_u^\dagger
(\tilde g_u \sigma\widetilde W + \tilde g'_u\widetilde B)\widetilde H_u
-\textstyle{\frac{1}{\sqrt{2}}} H_d^\dagger
(\tilde g_d \sigma \widetilde W - \tilde g'_d \widetilde B)\widetilde H_d 
+\mathrm{h.c.}
\nonumber
\end{eqnarray}
The first two lines correspond to the Higgs potential of a two Higgs doublet model,
where the quartic couplings have boundary conditions at $\widetilde m$ that connect
to the supersymmetric models above $\widetilde m$. In the third line we include the
Yukawa couplings, while in the forth one we have the Higgs-higgsino-gaugino
couplings. These $Y$ and $\tilde g$ couplings in PSS differ from the corresponding 
ones in SS in their RGE and their boundary conditions at $\widetilde m$.

BRpV is introduced in PSS with the terms,
\begin{equation}
{\cal L}_{PSS}^{RpV} =
\epsilon_i \widetilde H_u^T \epsilon L_i  \ -\ 
{\frac{1}{\sqrt{2}}} b_i H_u^T\epsilon
(\tilde g_d \sigma\widetilde W-\tilde g'_d\widetilde B)L_i 
\ + \ h.c., 
\label{LSS2HDMRpV}
\end{equation}
where the origin of the second term is analogous as in SS: they are generated as
effective terms after integrating out the heavy sfermions.

\subsection{Neutrinos and Neutralinos in PSS}
\label{subsec_neutral_PSS}

The neutralino sector of the neutrino/neutralino mass matrix in PSS
has the following form,
\begin{equation}
{\bf M}_{\chi^0}^{PSS}=\left(\begin{array}{cccc}
M_1 & 0 & -\frac{1}{2}\tilde g'_d v_d & \frac{1}{2}\tilde g'_u v_u \\
0 & M_2 & \frac{1}{2}\tilde g_d v_d & -\frac{1}{2}\tilde g_u v_u \\
-\frac{1}{2}\tilde g'_d v_d & \frac{1}{2}\tilde g_d v_d & 0 & -\mu \\
\frac{1}{2}\tilde g'_u v_u & -\frac{1}{2}\tilde g_u v_u & -\mu & 0
\end{array}\right).
\label{X0massmat2}
\end{equation}
It differs only slightly from SS in eq.~(\ref{X0massmat}):
it is apparent in eq.~(\ref{X0massmat2}) that there are two different vacuum 
expectation values, as in the MSSM, and as it was mentioned before the $\tilde g$ 
couplings have different RGE and boundary conditions. The mixing sub-matrix has
also only minor differences,
\begin{equation}
m^{PSS}=\left(\begin{array}{cccc}
-\frac{1}{2} \tilde g'_d b_1 v_u & 
 \frac{1}{2} \tilde g_d  b_1 v_u & 
0 &\epsilon_1 
\cr
-\frac{1}{2} \tilde g'_d b_2 v_u & 
 \frac{1}{2} \tilde g_d  b_2 v_u & 
0 & \epsilon_2 
\cr
-\frac{1}{2} \tilde g'_d b_3 v_u & 
 \frac{1}{2} \tilde g_d  b_3 v_u &
0 & \epsilon_3
\end{array}\right),
\end{equation}
The neutrino effective mass matrix in PSS takes the form,
\begin{equation}
{\bf M}_\nu^{eff}=
\frac{M_1 \tilde g^2_d + M_2 \tilde g'^2_d}{4\det{M_{\chi^0}^{PSS}}}
\left(\begin{array}{cccc}
\Lambda_1^2        & \Lambda_1\Lambda_2 & \Lambda_1\Lambda_3 \cr
\Lambda_2\Lambda_1 & \Lambda_2^2        & \Lambda_2\Lambda_3 \cr
\Lambda_3\Lambda_1 & \Lambda_3\Lambda_2 & \Lambda_3^2
\end{array}\right),
\label{treenumass2}
\end{equation}
with $\Lambda_i=\mu b_i v_u + \epsilon_i v_d$, and with the determinant 
of the neutralino submatrix equal to,
\begin{equation}
\det{M_{\chi^0}^{PSS}}=-\mu^2 M_1 M_2 + \frac{1}{2} v_uv_d\mu \left( 
M_1 \tilde g_u\tilde g_d + M_2 \tilde g'_u \tilde g'_d \right)
+\textstyle{\frac{1}{16}} v_u^2v_d^2 
\left(\tilde g'_u \tilde g_d - \tilde g_u \tilde g'_d \right)^2.
\label{detNeut2}
\end{equation}
Despite these differences, the tree-level neutrino mass matrix in 
eq.~(\ref{treenumass2}) also has only one non-zero eigenvalue, generating 
an atmospheric mass difference but not a solar mass difference.
Nevertheless, as oppose to the SS case, in PSS quantum corrections do
lift the symmetry of the tree-level matrix, generating a corrected
neutrino mass matrix that looks like,
\begin{equation}
M_\nu^{ij}=A\Lambda_i\Lambda_j+C\epsilon_i\epsilon_j,
\label{DpiH2HDM}
\end{equation}
where the tree-level value $A^{(0)}$ can be read from eq.~(\ref{treenumass2}).
One-loop diagrams corrects it into the value $A$, and generate the constant $C$.
The matrix in eq.~(\ref{DpiH2HDM}) has only one null eigenvalue, thus a
non-zero atmospheric and solar mass difference. A quadratic constant $B$ that 
mixes $\Lambda_i$ and $\epsilon_j$ is also generated in general, but can be 
adjusted to zero by choosing an appropriate value for the arbitrary 
renormalization scale of dimensional regularization.

This mechanism depends strongly on the $b_i$ terms
in the definition of $\Lambda_i$.
The origin of those terms is that
above the splitting scale $\tilde m$ the Higgs scalars gauge eigenstates mix with
sneutrinos gauge eigenstates. 
This happens for the CP-even real parts and the CP-odd imaginary parts.
Due to this mixing one might define the real part of sneutrinos ($s^i_s, \, t^i_s$) 
in the CP-even Higgs mass eigenstates ($h,\, H$).
It has been shown that for the real part
$s^i_s \sim -b_i c_\alpha \sim -c_\alpha v_i /v_u$ and that
$t^i_s\sim  -b_i s_\alpha \sim -s_\alpha v_i /v_u$.
The relations for the imaginary parts are analogous \cite{Diaz:2006ee}.
Thus, the existence of a non-zero $b_i$ term in indicates that actually 
Higgs Bosons, at any energy scale below $\tilde m$ will have a small sneutrino component.
This means that the sneutrinos are not completely decoupled at those scales.
It is further instructive to notice that the $b_i$
are proportional to the sneutrino vacuum expectation value, 
which implies that it disappears for a restored
$SU(2)$ symmetry. This fact is important in order to understand
this model in context of some general theorems on neutrino 
masses~\cite{Weinberg:1979sa,hep-ph/9701253}.

\subsection{Charged Leptons and Charginos in PSS}
\label{subsec_charged_PSS}

The chargino block in PSS has the following structure,
\begin{equation}
M_{\chi^+} = \left( \begin{matrix}
M_2 & \frac{1}{\sqrt{2}}\tilde g_u v_u \cr \frac{1}{\sqrt{2}}\tilde g_d v_d & \mu
\end{matrix} \right)
\end{equation}
The difference with the SS case in eq.~(\ref{chaSS}) lies in the fact that now
we have two vacuum expectation values $v_u$ and $v_d$ (as in the MSSM), and
that the $\tilde g$ couplings, defined in eq.~(\ref{LagSS2HDM}), are numerically
different.

The mixing between charginos and charged leptons is given by the matrices
\begin{equation}
m_c = \left( \begin{matrix}
\frac{1}{\sqrt{2}}\tilde g_d b_1 v_d & -\epsilon_1 \cr
\frac{1}{\sqrt{2}}\tilde g_d b_2 v_d & -\epsilon_2 \cr
\frac{1}{\sqrt{2}}\tilde g_d b_3 v_d & -\epsilon_3 \cr
\end{matrix} \right)
\end{equation}
and
\begin{equation}
Y = \left( \begin{matrix}
0 & 0 & 0 \cr
-\frac{1}{\sqrt{2}}Y'^{1i}_\ell b_i v_d & -\frac{1}{\sqrt{2}}Y'^{2i}_\ell b_i v_d &
-\frac{1}{\sqrt{2}}Y'^{3i}_\ell b_i v_d  \cr
\end{matrix} \right)
\end{equation}
where $Y'$ is the charged lepton Yukawa matrix in our second ansatz. The dimensionless 
parameter $b_i$ plays the same role as $a_i$ in SS.

As in SS, in this scenario we consider the charged lepton Yukawa coupling matrix 
as non-diagonal, and we consider its effect in the relation between neutrino 
parameters and observables.

\subsection{Effects on Neutrino Parameters in PSS}
\label{subsec_parameters_PSS}

The neutrino mass matrix in PSS is given by eq.~(\ref{DpiH2HDM}), and in 
what we call tree-level dominance scenario, defined by 
$A^2|\vec\Lambda|^4 \gg C^2|\vec\epsilon\,|^4$, the neutrino mass differences are
found to be,
\begin{equation}
\Delta m^2_{atm} \approx A^2|\vec\Lambda|^4
\,,\qquad
\Delta m^2_{sol} \approx 
C^2 \frac{|\vec\epsilon\times\vec\Lambda|^4}{|\vec\Lambda|^4}\quad.
\end{equation}
These expressions are not changed by the presence of a non-trivial diagonalization
matrix $V_L$ for the charged lepton mass matrix.

The three normalized eigenvectors of the neutrino mass matrix in eq.~(\ref{DpiH2HDM})
in the tree-level dominance scenario are, in first approximation,
\begin{equation}
\vec e_1 = \frac{\vec\epsilon\times\vec\Lambda}
{|\vec\epsilon\times\vec\Lambda|} \,,\qquad
\vec e_2 = \frac{\vec\Lambda\times(\vec\epsilon\times\vec\Lambda)}
{|\vec\Lambda\times(\vec\epsilon\times\vec\Lambda)|} \,,\qquad
\vec e_3 = \frac{\vec\Lambda}{|\vec\Lambda|}
\end{equation}
and they form the columns of the $U_\nu$ matrix. The neutrino mixing angles
written in terms of the approximated mixing angles (when $V_L=1$) are 
displayed in eq.~(\ref{anlgeschange}), while their expressions written in 
terms of the BRpV parameters [analogous to eq.~(\ref{anglesSS})] are more 
involved, and we display them in terms of the eigenvector components, and
in the approximation $\sin\theta', \sin\alpha' \ll 1$,
\begin{eqnarray}
\sin\theta_{13} &=& e_{31} + s_{\alpha'} e_{32}
\nonumber\\
\tan\theta_{23} &=& \frac{e_{32}}{e_{33}} \bigg[
1+s_{\theta'} \Big( \frac{e_{32}}{e_{33}}+\frac{e_{33}}{e_{32}} \Big)
-s_{\alpha'} \frac{e_{31}}{e_{32}} \bigg]
\label{anglesPSS}\\
\tan\theta_{12} &=& \frac{e_{21}}{e_{11}} \bigg[
1+s_{\alpha'} \Big( \frac{e_{22}}{e_{21}}-\frac{e_{12}}{e_{11}} \Big)
\bigg] \nonumber
\end{eqnarray}
where $e_{ij}$ refers to the component $j$ of the eigenvector $\vec e_i$.
The numerical effect will be shown in the next section.

\section{Numerical Results}
\label{sec_Numerical}

In this analysis, prediction of neutrino parameters is done by using
numerical methods to find the eigenvalues and eigenvectors that correspond
to $U_{\nu}$ and $V_L$. Using them, we find neutrino mass differences
and mixing angles, and compare them with values from experimental measurements. 
We also study how a non-diagonal Yukawa matrix can influence the neutrino 
observables, specifically neutrino mixing angles.

The agreement with the experimental boundaries at the
$3\sigma$-level was quantified by
calculating \cite{M:2004,Schwetz:2011qt,Schwetz:2011zk}
\begin{align}
\chi^{2}=&\left(\displaystyle{\frac{10^{3}\Delta m^{2}_{atm} - 
2,45}{0,31}}\right)^{2} +
\left(\displaystyle{\frac{10^{5}\Delta m^{2}_{sol} - 7,64}{0,55}}\right)^{2}
\nonumber\\  & +
 \left(\displaystyle{\frac{\sin^2{\theta_{atm}}-0,515}{0,125}}\right)^{2} + 
 \left(\displaystyle{\frac{\sin^2{\theta_{sol}}-0,315}{0,045}}\right)^{2}+
\left(\displaystyle{\frac{\sin^2{\theta_{rea}}-0,018}{0,017}}\right)^{2},
\label{x2}
\end{align}
where $\theta_{atm}=\theta_{23}$, $\theta_{sol}=\theta_{12}$, and
$\theta_{reac}=\theta_{13}$.
We accept values of $\chi^2<1$ to be consistent with experimental results.

\subsection{Split SUSY}
\label{subsec_numerical_SS}

Our parameter space in SS can be classified into four type of variables. 
First supersymmetric parameters like the bino mass $M_{1}$, the wino mass 
$M_{2}$, the Higgsino mass $\mu$, and the ratio between vevs $\tan{\beta}$, 
whose effect can be all concentrated into the parameter $A_\lambda$ defined in 
eq.~(\ref{eq_mixmod}). Second the BRpV parameters $\lambda_{i}$,
which give rise to an atmospheric mass. Third the gravity parameter 
$\mu_{g}$ responsible for a solar mass. Fourth the charged
lepton GUT parameters $E$, $F$, $D$ which define the angle $\alpha$ 
in eq.~(\ref{GJ}).

We scan the parameter space varying randomly $A_\lambda$, the BRpV parameters 
$\lambda_i$, and the gravity parameter $\mu_g$, looking for a solution
with good prediction for the neutrino parameters. In order to compare 
easily with PSS we define $\Lambda_i=v_d\lambda_i$ 
and $A=A_\lambda/v_d^2$ for the SS case.
As a working point we choose the numerical values given in Table \ref{t1}, 
with the values of $M_1$, $M_2$, $\mu$, and $\tan\beta$ as an example 
of a set that leads to the corresponding value for $A$. The charged 
lepton GUT parameters are fixed to their values inferred by the 
Georgi-Jarlskog ansatz in eq.~(\ref{GJansatz}), which lead to 
$\sin{\alpha}\approx 0.0695$. This solution is in good agreement with 
all neutrino observables, with the predictions shown in Table \ref{t11}.
%
\begin{table}[h]
\scriptsize
\begin{minipage}{10cm}
\begin{tabular}{|c|c|c|c|} \hline\hline
SUSY Parameters & Value & Scanned Range & Units\\ \hline\hline
$M_{1}$              & $177$  & $[40,500]$ & $\mbox{GeV}$ \\ \hline
$M_{2}$              & $300$  & $[80,100]$ & $\mbox{GeV}$ \\ \hline
$\left|\mu\right|$   & $392$  & $[0,1000]$ & $\mbox{GeV}$ \\ \hline
$\tan{\beta}$        & $25.1$   & $[2,50]$   & -       \\ \hline\hline
$A$                  & $-3.53$   & $-$   &  eV/GeV$^{4}$   \\ \hline\hline
BRpV Parameters      &  Value & Scanned Range & Units\\ \hline\hline
$\Lambda_{1}$        & $0.0109$ &  $[-1,1]$         & $\mbox{GeV}^2$ \\ \hline
$\Lambda_{2}$        & $-0.0873$  &  $[-1,1]$         &  $\mbox{GeV}^2$ \\ \hline
$\Lambda_{3}$        & $0.0814$ &  $[-1,1]$         &  $\mbox{GeV}^2$ \\ \hline 
Gravity Parameter & Value & Scanned Range& Units\\ \hline\hline
$\mu_{g}$ & $0.00291$ &$[0,0.005]$ &eV \\ \hline
\end{tabular}
\caption{\it Solutions for the parameters. This values gives $\chi^{2}=0.356$.}
\label{t1}
\end{minipage}
\ \
\hfill \begin{minipage}{5cm}
\begin{tabular}{|c|c|c|} \hline
 Observable       & Solution & Units \\ \hline\hline
$\Delta m_{atm}^{2}$      & $2.44\times10^{-3}$  & $\mbox{eV}^{2}$ \\ \hline
$\Delta m_{sol}^{2}$      & $7.61\times10^{-5}$  & $\mbox{eV}^{2}$ \\ \hline
$\sin^{2}{\theta}_{atm}$  & $0.532$              & -      \\ \hline
$\sin^{2}{\theta}_{sol}$  & $0.290$              & -       \\ \hline
$\sin^{2}{\theta}_{rea}$  &  $0.0195$              & -       \\ \hline
\end{tabular}
\caption{\it SS predictions for neutrino observables given the values in 
Table \ref{t1}, and $\sin{\alpha}\approx 0.0695$.}
\label{t11}
\end{minipage}
\end{table}

\begin{figure}[h]
\begin{minipage}{8cm}
\centerline{\protect\vbox{\epsfig{file=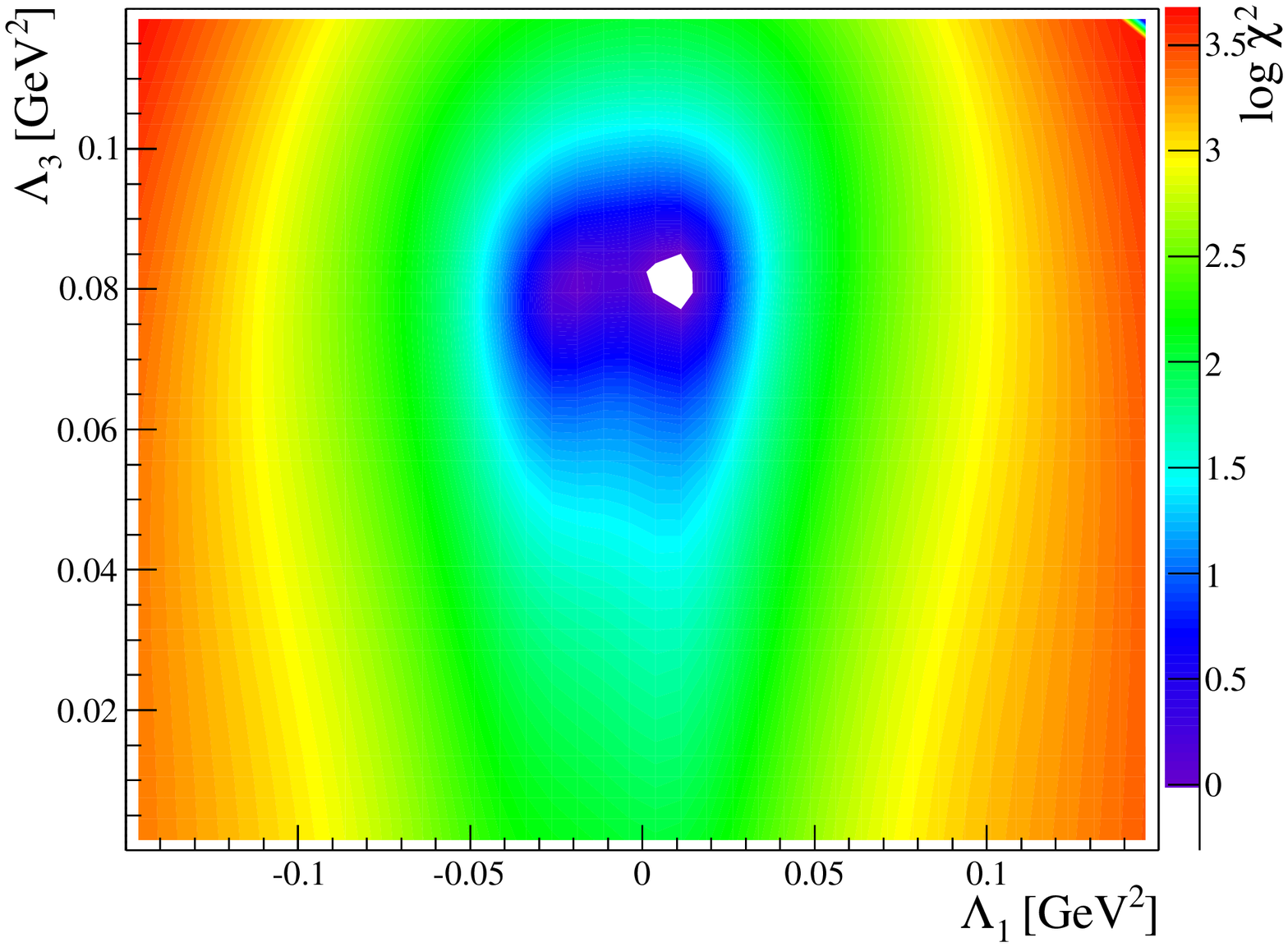,width=1\textwidth}}}
\end{minipage}
\ \
\hfill \begin{minipage}{8cm}
\centerline{\protect\vbox{\epsfig{file=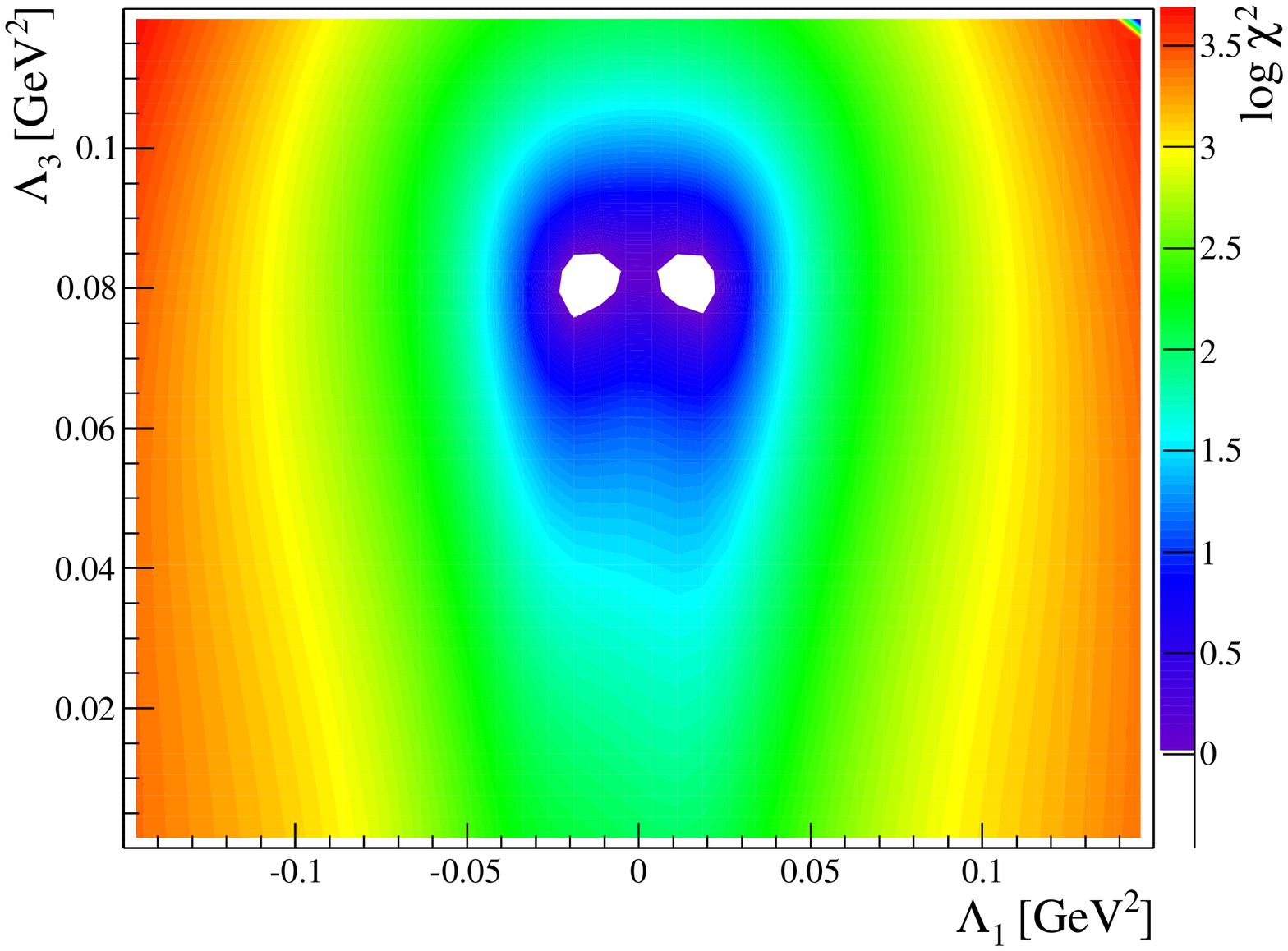,width=1\textwidth}}}
\end{minipage}
\caption{\it $\chi^{2}$ as a function of the BRpV parameters $\Lambda_{1}$ 
and $\Lambda_{3}$ for $\sin{\alpha}\approx 0.0695$ (left) and $\sin{\alpha}=0$ 
(right), keeping the rest of the parameters as indicated in Table \ref{t1}.}
\label{xi2all}
\end{figure}

In Fig.~\ref{xi2all}-left we plot the logarithm of $\chi^{2}$ as 
contour regions in the $\Lambda_{1}$ and $\Lambda_{3}$ plane, with fixed values for 
all the other parameters as indicated in Table \ref{t1}, plus
$\sin{\alpha}\approx 0.0695$.
%
%
Good solutions to neutrino observables are represented by the white region, 
corresponding to $\chi^{2}<1$. We see that the countours are not symmetric 
under a $\Lambda_1$ sign change. This is due to the $\chi^2$-term 
corresponding to the reactor angle in eq.~(\ref{x2}), and can be understood 
from eq.~(\ref{anglesSS}). We see that the correction to the reactor angle 
due to a non diagonal charged lepton mass matrix is large 
because, in addition to the fact that $\sin\alpha\ll\cos\alpha$, we also have
$|\Lambda_2|\gg|\Lambda_1|$ compensating the previous disbalance. Therefore,
$\sin\theta_{13}$ is not symmetric under a change in the sign of $\Lambda_1$
unless it is accompanied by a corresponding change in the sign of $\Lambda_2$.

In order to see the effect of the diagonalization of the charged lepton mass
matrix, we compare the same effect as before but now setting $\sin\alpha=0$, 
which is equivalent to a diagonal charged lepton mass matrix. This is done
in Fig.~\ref{xi2all}-right, where we have the analogous countour plot for 
$\chi^2$. One sees that for the chosen point in parameter space, the allowed 
(white) region for the case $\sin{\alpha}\approx 0.0695$ (Fig.~\ref{xi2all}-left) 
is smaller than the corresponding region for the case $\sin{\alpha}=0$ 
(Fig.~\ref{xi2all}-right). This means that points in parameter space consistent 
with neutrino observables when the diagonalization of the charged lepton mass 
matrix is neglected, can actually be inconsistent when this diagonalization is 
taken into account. In addition, an approximated symmetry under the $\Lambda_1$ 
sign change is re-established in the case of $\sin{\alpha}=0$. This 
is because in this case $\sin^2\theta_{rea}$ is insensitive to this sign.

The previous conclusions are confirmed when we study separately the effect on 
$\chi^2$ from the neutrino angles. We remind the reader that the neutrino masses
are not affected by the diagonalization matrix in the charged lepton sector,
as we explained below eq.~(\ref{eq_mixmod}). In addition, the effect of the 
non-diagonal charged lepton matrix on the atmospheric angle is relatively small. 
The solar and reactor angles however get significant changes after the 
inclusion of charged lepton diagonalization effects. To show this we define,
\begin{equation}
\chi^{2}_{s2sol}=
\left(\displaystyle{\frac{\sin^2{\theta_{sol}}-0,33}{0,07}}\right)^{2}
\,,\qquad
\chi^{2}_{s2rea}=
\left(\displaystyle{\frac{\sin^2{\theta_{rea}}-0,018}{0,017}}\right)^{2}
\label{s2solyrea}
\end{equation}
which are the isolated contributions to $\chi^2$ from the solar and reactor angles
respectively.
%
\begin{figure}[h]
\begin{minipage}{8cm}
\centerline{\protect\vbox{\epsfig{file=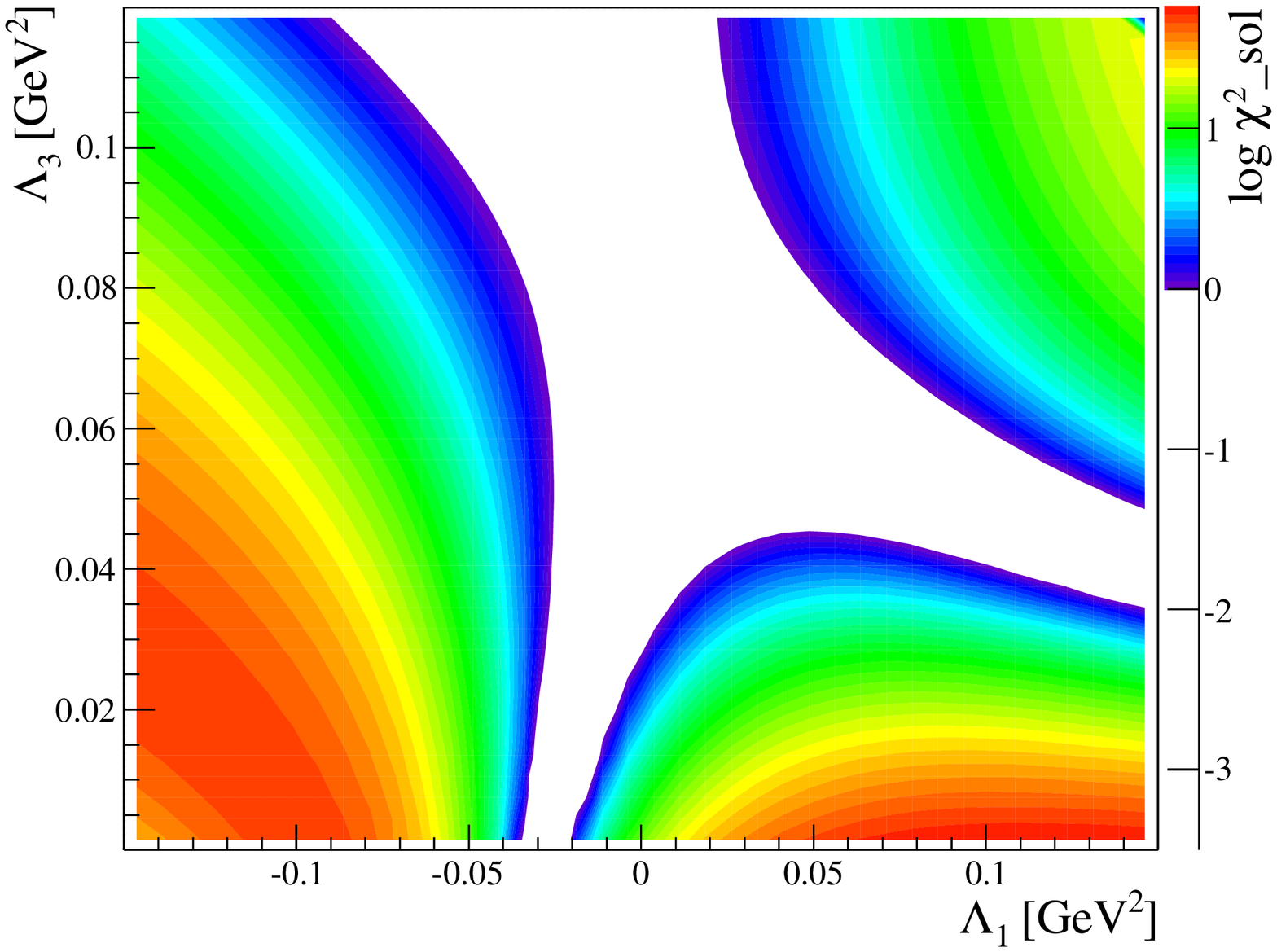,width=1\textwidth}}}
\end{minipage}
\ \
\hfill \begin{minipage}{8cm}
\centerline{\protect\vbox{\epsfig{file=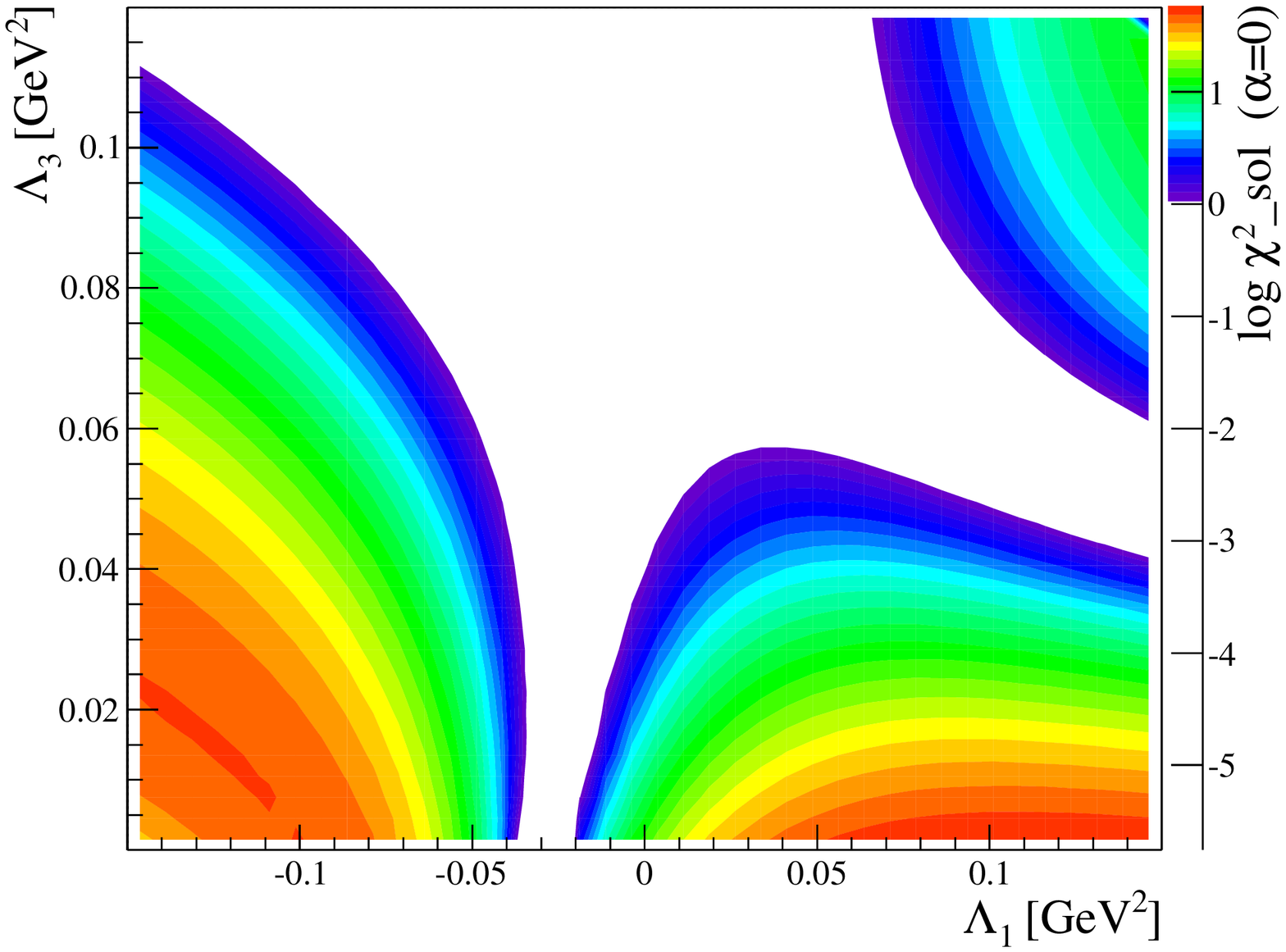,width=1\textwidth}}}
\end{minipage}
\caption{\it $\chi^{2}_{s2sol}$ as a function of the BRpV parameters $\Lambda_{1}$ 
and $\Lambda_{3}$ for $\sin{\alpha}\approx 0.0695$ (left) and $\sin{\alpha}=0$ 
(right), keeping the rest of the parameters as indicated in Table \ref{t1}.}
\label{Xi2sol}
\end{figure}
%
In Fig.~\ref{Xi2sol} we have $\chi^{2}_{s2sol}$, with $\sin{\alpha}\approx 0.0695$
in the left frame and $\sin{\alpha}=0$ in the right one. We see important differences
in the shape of the allowed region (white). Nevertheless the overall significance is 
decreased because the contribution from the solar angle to $\chi^2$ is relatively 
small.
%
\begin{figure}[h]
\begin{minipage}{8cm}
\centerline{\protect\vbox{\epsfig{file=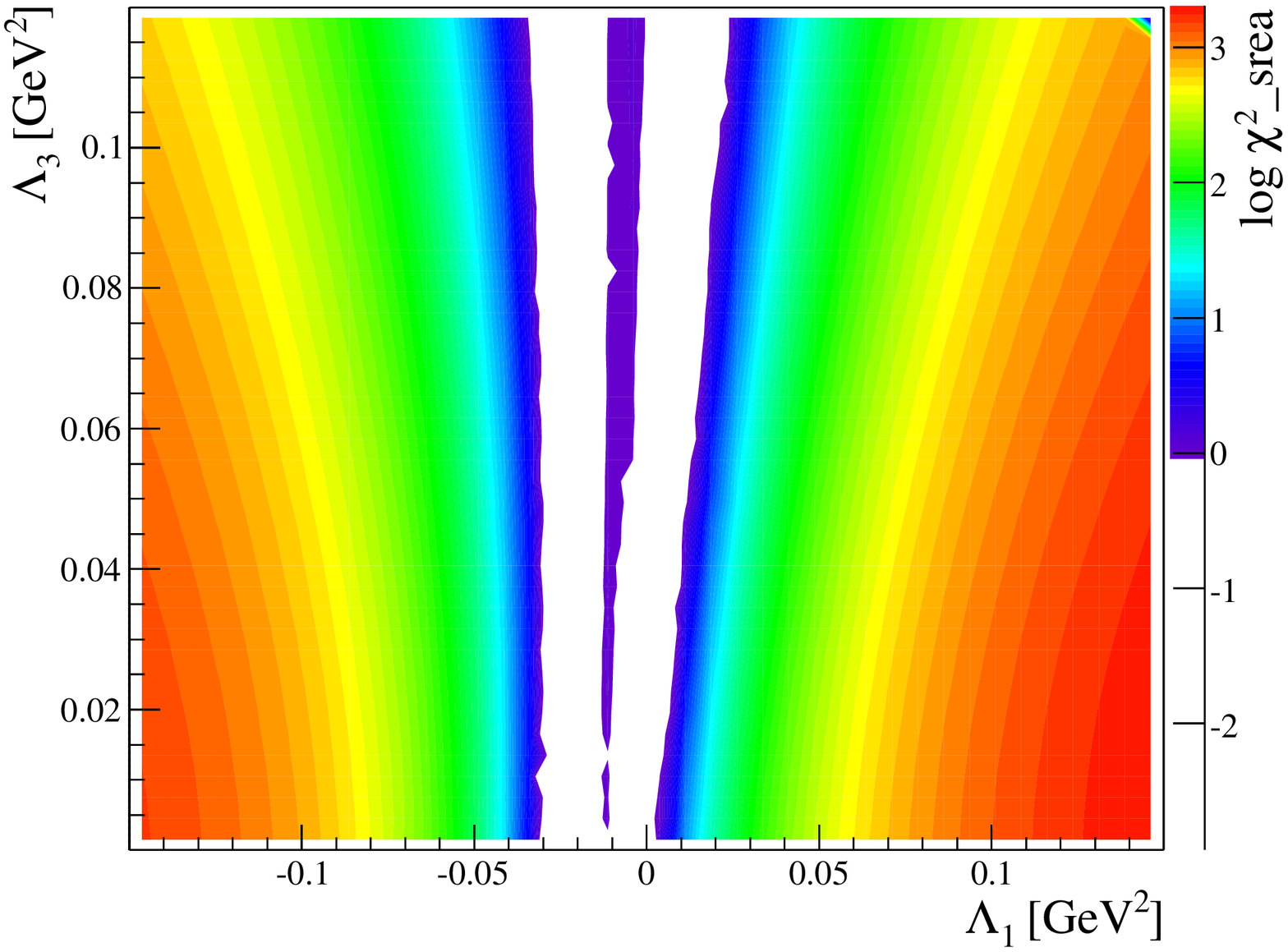,width=1\textwidth}}}
\end{minipage}
\ \
\hfill \begin{minipage}{8cm}
\centerline{\protect\vbox{\epsfig{file=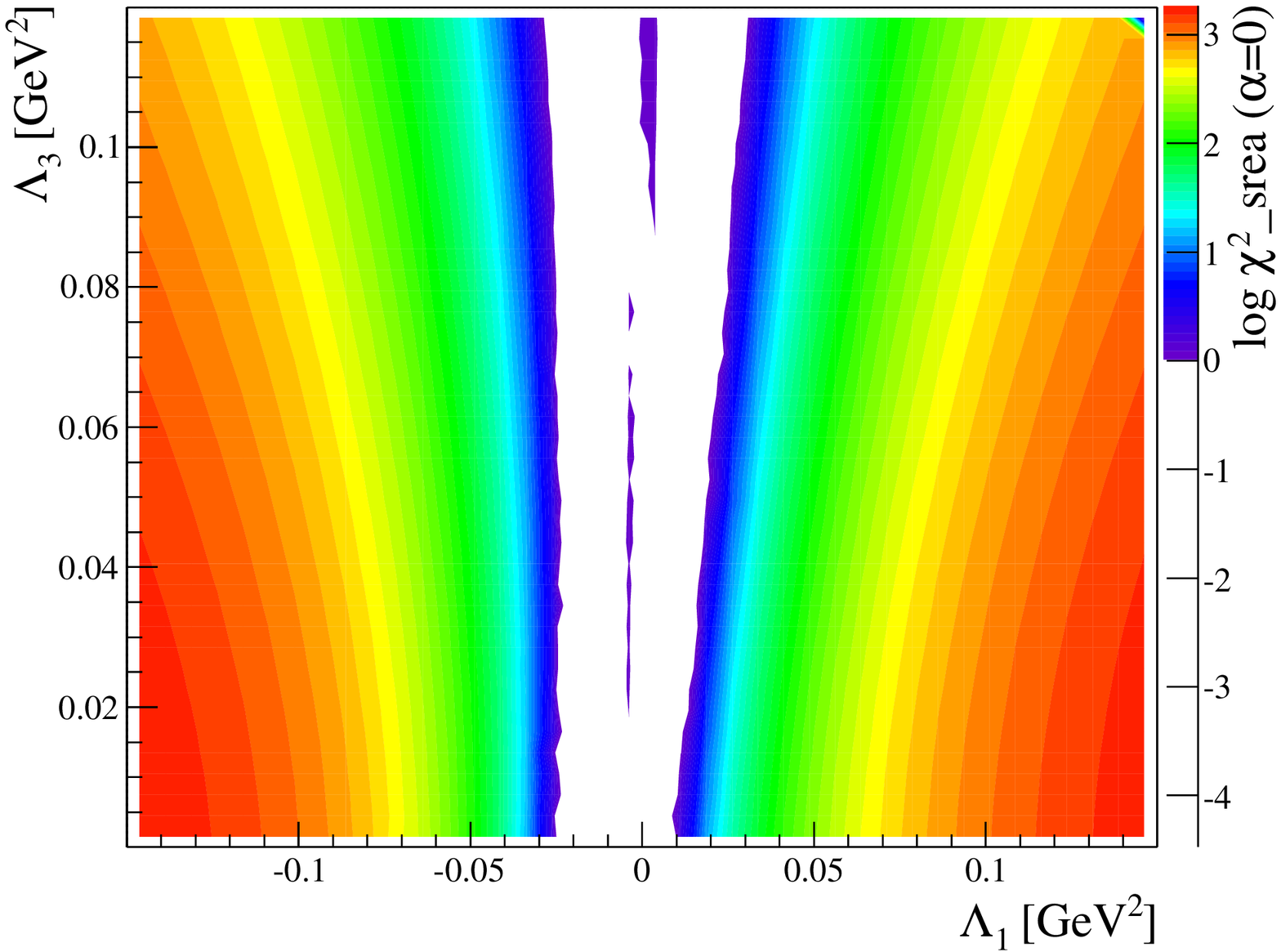,width=1\textwidth}}}
\end{minipage}
\caption{\it $\chi^{2}_{s2rea}$ as a function of the BRpV parameters $\Lambda_{1}$ 
and $\Lambda_{3}$ for $\sin{\alpha}\approx 0.0695$ (left) and $\sin{\alpha}=0$
(right), keeping the rest of the parameters as indicated in Table \ref{t1}.}
\label{Xi2rea}
\end{figure}
%
On the other hand, in Fig.~\ref{Xi2rea} we have $\chi^{2}_{s2rea}$ with an 
analogous difference between left and right frames. The shift in the allowed 
region from left ($\sin{\alpha}\approx 0.0695$) to right ($\sin{\alpha}=0$) 
is much smaller than in the solar angle case, but the numerical contribution
to $\chi^2$ from the reactor angle is much larger, making the reactor
angle the most decisive factor in the influence of the diagonalization
of the charged lepton mass matrix. We also mention that the prediction in 
\cite{Diaz:2009yz} that $\mu_g= {\mathcal{O}}(0.01)$~eV is not affected by 
the scenario where the charged lepton mass matrix is not diagonal, since 
$\mu_g$ is in first approximation restricted only by mass differences.

\subsection{Partial Split SUSY}
\label{subsec_numerical_PSS}

In PSS the parameter space consists of, first, the supersymmetric parameters 
Bino mass $M_{1}$, Wino mass $M_{2}$, Higgsino mass $\mu$, $\tan{\beta}$,
and Higgs masses $m_h$ and $m_A$, which define the constants $A$ and $C$ 
in eq.~(\ref{DpiH2HDM}); second, the BRpV parameters $\Lambda_{i}$ and 
$\epsilon_{i}$; and third, the charged lepton Yukawa parameters $E$, $F$, 
and $D$, which define the angles $\sin{\alpha}'$ and $\sin{\theta}'$ in 
(\ref{Giudice}).

As we did for the previous model, we perform a scan over parameter space and look
for solutions with predictions on neutrino observables compatible with 
experimental data, represented by the value of $\chi^2<1$ as given in 
eq.~(\ref{x2}). A working scenario satisfying this criteria is given in
Table~\ref{t2}. The effect of the first 6 parameters is in the values of
$A$ and $C$ which enter in the neutrino mass matrix. The scale $Q$ is chosen 
such that there is no mixing term between $\Lambda$ and $\epsilon$. The
scenario is completed with the values of the BRpV parameters $\Lambda_i$
and $\epsilon_i$. In Table~\ref{t22} we have the predictions for the neutrino 
observables in this model, which gives a value of $\chi^2=0.88$.
%
\begin{table}[h]
\scriptsize
\begin{minipage}{10cm}
\begin{tabular}{|c|c|c|c|} \hline\hline
SUSY Parameters & Value & Scanned Range & Units\\ \hline\hline
$M_{1}$            & $119$     & $[40,500]$    & $\mbox{GeV}$ \\ \hline
$M_{2}$            & $339$     & $[80,100]$    & $\mbox{GeV}$ \\ \hline
$|\mu|$ & $456$     & $[0,1000]$    & $\mbox{GeV}$ \\ \hline
$\tan{\beta}$      & $5.71$      & $[2,50]$      & -            \\ \hline
$m_{h}$            & $130$     & $[114,140]$   & $\mbox{GeV}$ \\ \hline
$m_{A}$            & $1963$    & $[500,6000]$  & $\mbox{GeV}$ \\ \hline\hline
$A$    & $-2.73$     & -             & ${\mathrm{eV}}/{\mathrm{GeV}}^4$ \\ \hline
$C$    & $0.282$     & -             & ${\mathrm{eV}}/{\mathrm{GeV}}^2$ \\ \hline
$Q$                & $1048$     & -             & $\mbox{GeV}$ \\ \hline\hline
BRpV Parameters    &  Value    & Scanned Range & Units        \\ \hline\hline
$\Lambda_{1}$      & $0.0317$  &  $[-1,1]$     & $\mbox{GeV}^{2}$  \\ \hline
$\Lambda_{2}$      & $-0.0022$ &  $[-1,1]$     &  $\mbox{GeV}^{2}$ \\ \hline
$\Lambda_{3}$      & $0.0738$  &  $[-1,1]$     &  $\mbox{GeV}^{2}$ \\ \hline 
$\epsilon_{1}$     & $0.034$  &  $[-1,1]$     &  $\mbox{GeV}$     \\ \hline
$\epsilon_{2}$     & $0.264$  &  $[-1,1]$     &  $\mbox{GeV}$     \\ \hline
$\epsilon_{3}$     & $0.372$  &  $[-1,1]$     &  $\mbox{GeV}$     \\ \hline
\end{tabular}
\caption{\it Chosen values for PSS. This values gives $\chi^{2}=0.88$.}
\label{t2}
\end{minipage}
\ \
\hfill \begin{minipage}{5cm}
\begin{tabular}{|c|c|c|} \hline
 Observable       & Solution & Units \\ \hline\hline
$\Delta m_{atm}^{2}$      & $2.43\times10^{-3}$  & $\mbox{eV}^{2}$ \\ \hline
$\Delta m_{sol}^{2}$      & $7.66\times10^{-5}$  & $\mbox{eV}^{2}$ \\ \hline
$\sin^{2}{\theta}_{atm}$  & $0.495$              & -      \\ \hline
$\sin^{2}{\theta}_{sol}$  & $0.323$              & -       \\ \hline
$\sin^{2}{\theta}_{rea}$  & $0.0026$              & -       \\ \hline
\end{tabular}
\caption{\it PSS predictions for neutrino observables given the values in Table \ref{t2}.}
\label{t22}
\end{minipage}
\end{table}
%

\begin{figure}[htbp]
  \centering
  \begin{minipage}[b]{7 cm}
\epsfig{file=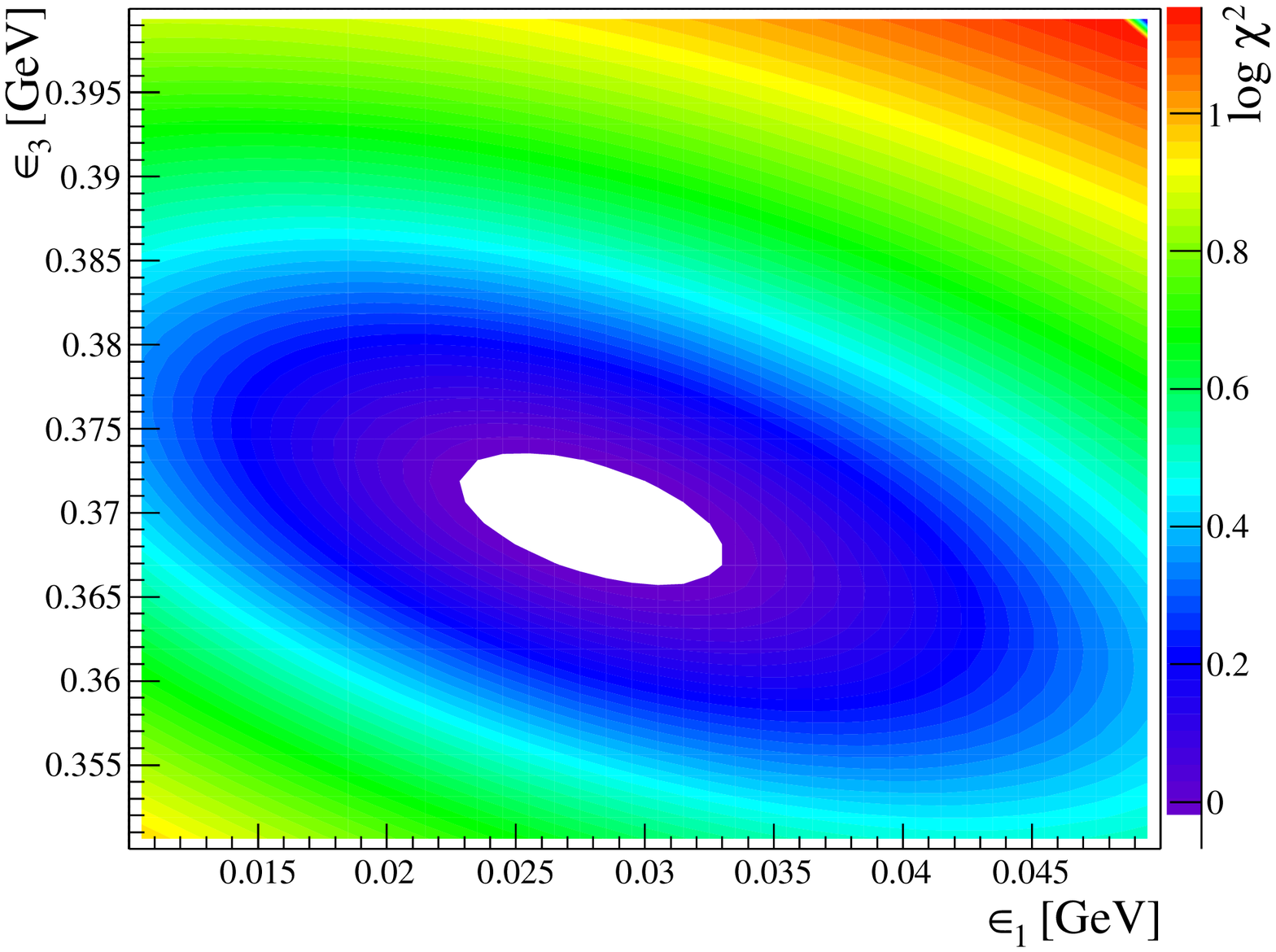,width=1.0\textwidth}
  \end{minipage}
\hspace{0.5 cm}
  \begin{minipage}[b]{7 cm}
\centerline{\protect\vbox{\epsfig{file=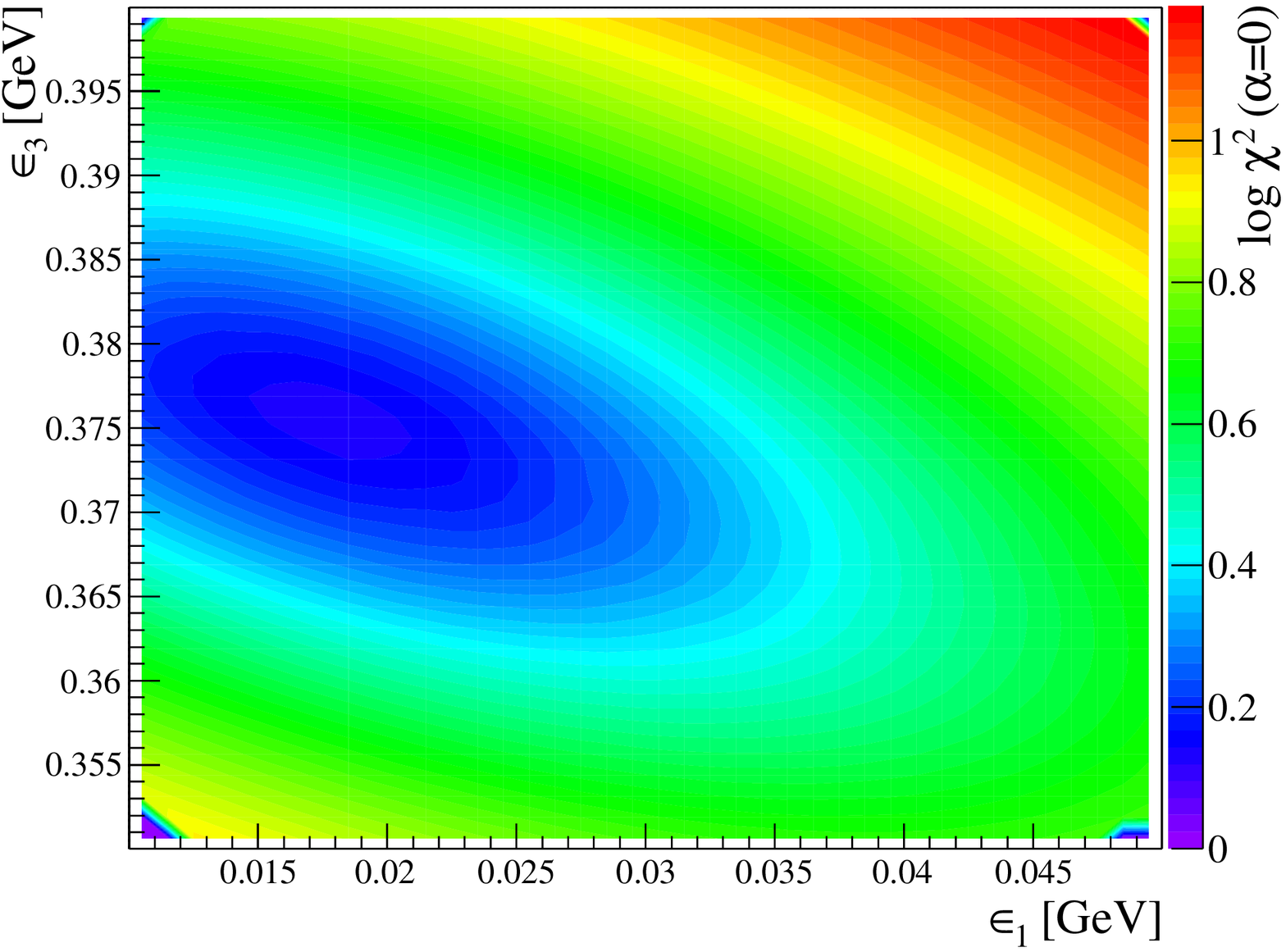,width=1.0\textwidth}}}
  \end{minipage}
\caption{\it $\chi^2$ in dependence of  $\epsilon_2$ and $\epsilon_3$,
while the other parameters are fixed around the central value from table\ref{t2}.
On the left hand side $\sin\alpha'=0.070$ and $\sin\theta'=0.036$ was used,
while on the right hand side a diagonal charged lepton matrix was used.}
\label{eps2eps3Diag}
\end{figure}

Similarly to the previous model, in Fig.~\ref{eps2eps3Diag}-left we have the 
logarithm of $\chi^2$ as contour regions in the $\epsilon_3$-$\epsilon_1$ plane, 
with all the other parameters fixed at their values in Table~\ref{t2}, plus
$\sin\alpha'=0.070$ and $\sin\theta'=0.036$. The white region corresponds to 
$\chi^2<1$, {\sl{i.e.}} points that satisfy the experimental constraints. 
Neglecting the effects of the diagonalization of the charged lepton mass matrix 
corresponds to set $\sin\alpha'=\sin\theta'=0$, and when this is done we find
$\chi^2=2.81$, meaning that a good point could have been missed if the charged 
lepton mass matrix diagonalization had not been taken into account. This can be 
seen graphically from Fig.~\ref{eps2eps3Diag}-right which is the analogous to 
the previous figure but neglecting the charged lepton mass matrix diagonalization.
%
%

\begin{figure}[ht]
\begin{minipage}{7cm}
\centerline{\protect\vbox{\epsfig{file=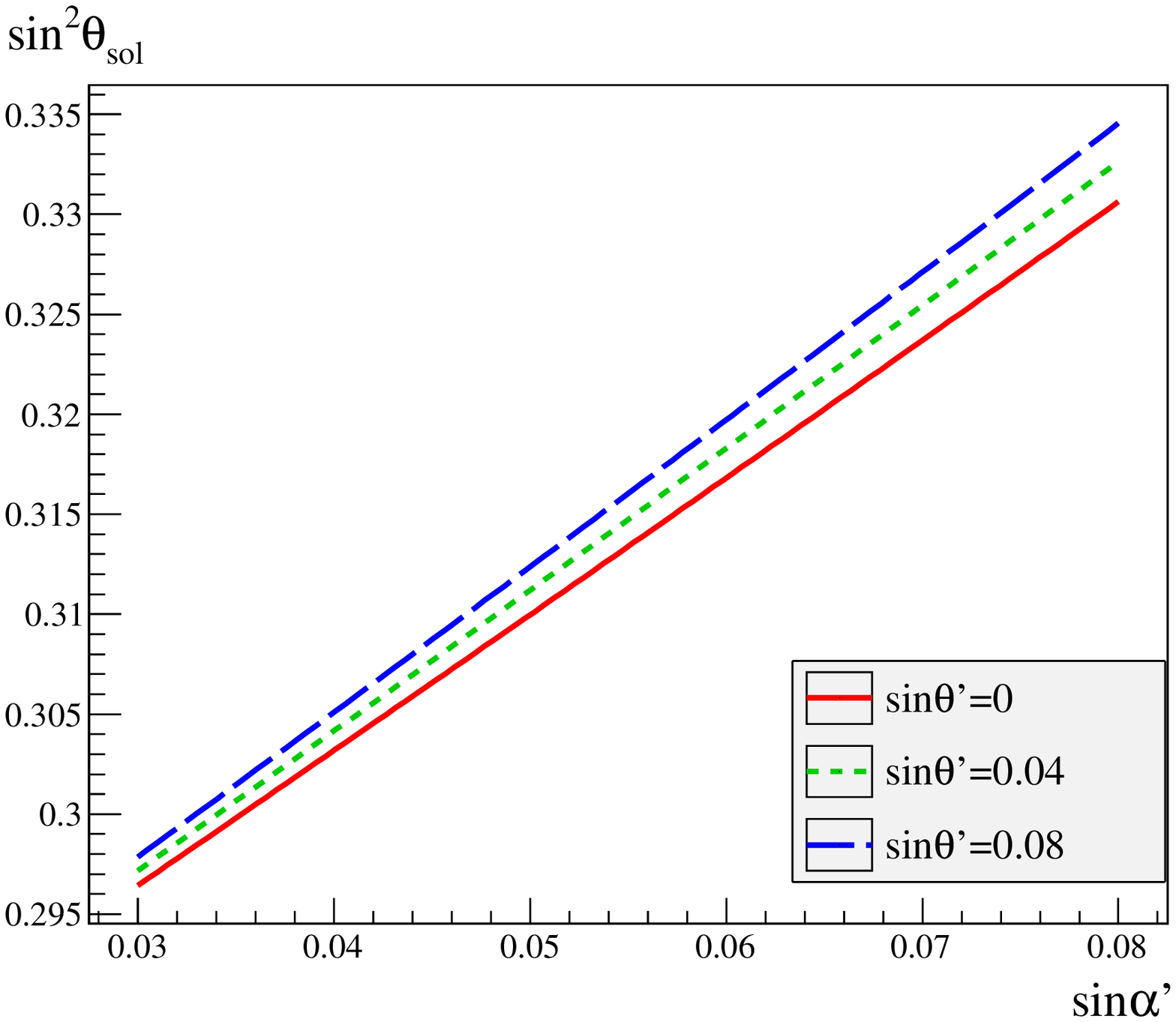,width=1\textwidth}}}
\end{minipage}
\ \
\begin{minipage}{7cm}
\centerline{\protect\vbox{\epsfig{file=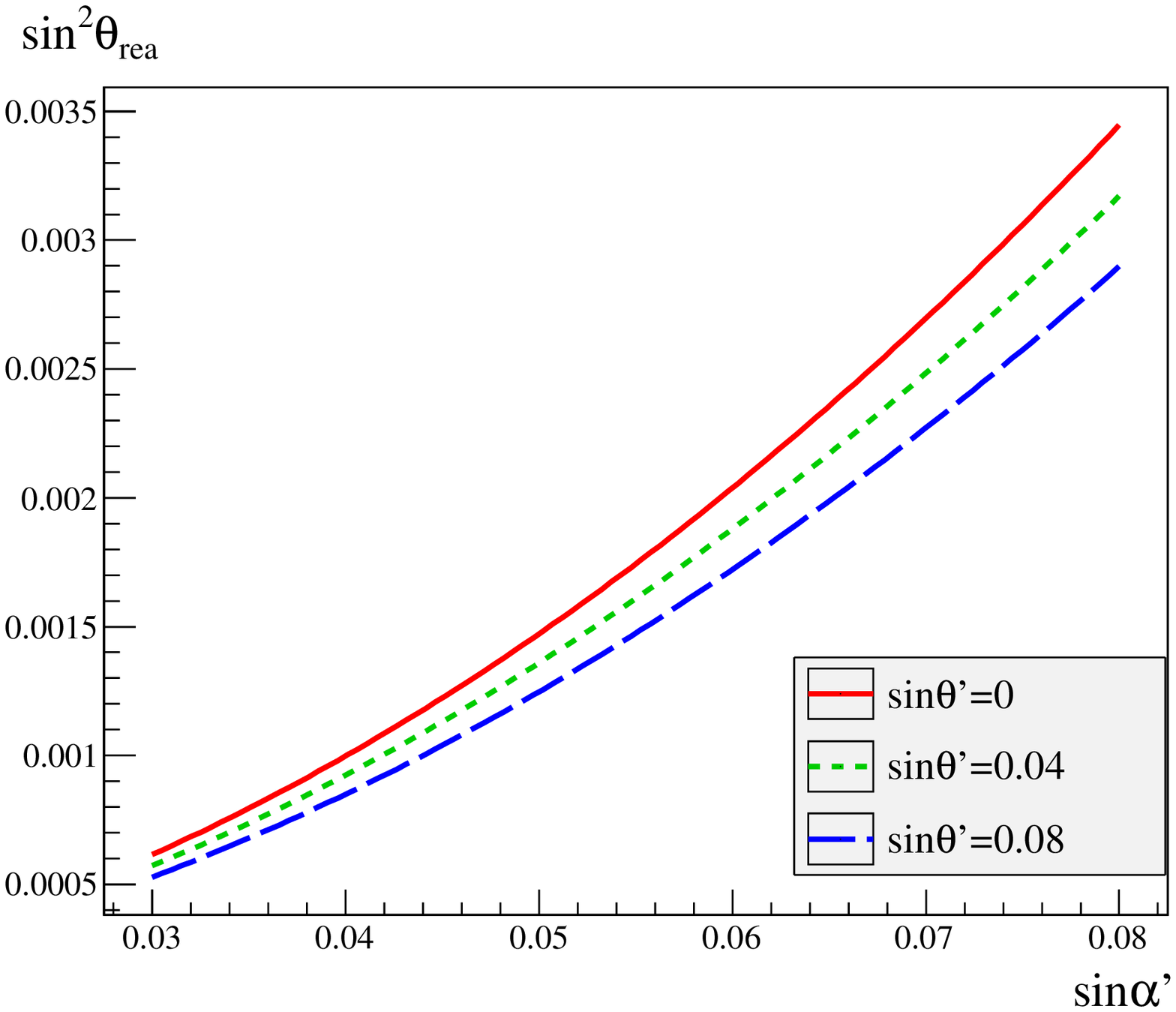,width=1\textwidth}}}
\end{minipage}
\caption{\it $\sin^2{\theta}_{sol}$ and $\sin^2{\theta}_{rea}$ dependence on 
$\sin{\alpha}'$ for different values of $\sin{\theta}'$. The other parameters are 
fixed around the central values in table \ref{t2}.}
\label{PPS1}
\end{figure}

It is useful to study the individual dependence of the neutrino angles on the
charged lepton rotation matrix angles $\alpha'$ and $\theta'$. In 
Fig.~\ref{PPS1} we have solar angle (left) and reactor angle (right) as a function
of $\sin\alpha'$ for three different values of $\sin\theta'$.
%
%
In both cases the dependence on $\sin\alpha'$ is stronger that the dependence on 
$\sin\theta'$, as can be noticed from eqs.~(\ref{anglesPSS}), where we see that 
the solar and reactor angles depend at first order only on $\sin\alpha'$, and a 
dependency on $\sin\theta'$ appears only at second order. Although the dependency
of the solar angle on $\sin\alpha'$ is strong, it variation on the chosen range
for $\sin\alpha'$ maintains the solar angle within its $3\sigma$ experimental 
region. On the contrary, the reactor angle being also very sensitive to $\sin\alpha'$
it can escape from below the experimental widow, while keeping its value well below 
the upper $3\sigma$ bound. Therefore, a lower bound on the reactor angle already 
constraints the model.

\begin{figure}[ht]
\begin{minipage}{7cm}
\centerline{\protect\vbox{\epsfig{file=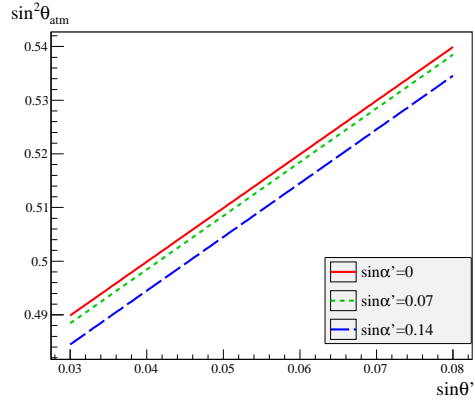,width=1\textwidth}}}
\end{minipage}
\caption{\it $\sin^2{\theta}_{atm}$ dependence on $\sin{\theta}'$ for different
values of $\sin{\alpha}'$. The other parameters are fixed around the central
values in table \ref{t2}.}
\label{PPS2}
\end{figure}

In Fig.~\ref{PPS2} we have a similar plot for the dependence of the atmospheric 
angle on $\sin{\theta}'$ for three different values of $\sin{\alpha}'$. As opposed
to the previous cases, for the atmospheric angle the dependence is stronger on 
$\sin{\theta}'$ rather than on $\sin{\alpha}'$. 
%
%
From eq.~(\ref{anglesPSS}) we see that despite the fact that $\tan\theta_{23}$ 
depends at first order on both angles, $\sin{\alpha}'$ is multiplied by the 
reactor angle and makes its influence much smaller. In any case, over the chosen 
range for $\sin{\theta}'$, the atmospheric angle does not leave the $3\sigma$ 
experimental window.

\begin{figure}[ht]
\centerline{\protect\vbox{\epsfig{file=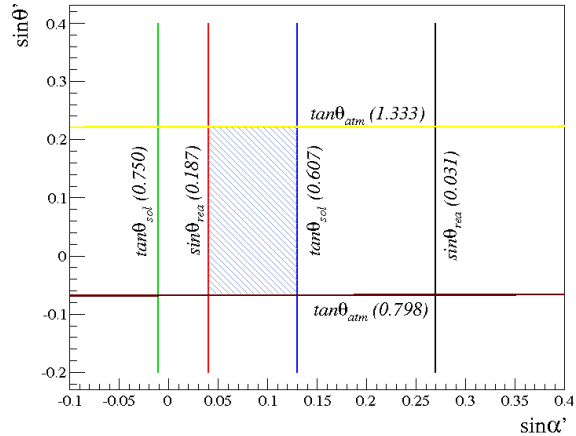,width=.5\textwidth}}}
\caption{\it Range of $\sin \theta'$ and $\sin \alpha'$
that gives $\chi^2<1$, for the scenario given in Table \ref{t2}.
}
\label{PPS3}
\end{figure}

In a related numerical analysis we plot in Fig.~\ref{PPS3} the allowed region 
(defined by $\chi^2<1$) in the $\sin{\theta}'$-$\sin{\alpha}'$ plane, with 
the effect of the different neutrino angle $3\sigma$ bounds shown as 
solid lines.
%
%
Here we confirm that the atmospheric angle restricts the values of $\sin{\theta}'$,
while the solar and reactor angles restricts the values of $\sin{\alpha}'$. The 
typical value for the charged lepton mixing angles in the Giudice ansatz are
$\sin{\theta}'=0.036$ and $\sin{\alpha}'=0.07$, and $\theta'$ will start to be probed 
if the error in the atmospheric angle diminishes by a few times. On the other hand
the value of $\alpha'$ can be probed with an improvement on the lower bound
of the reactor angle, and with an improvement on the upper bound of the solar angle.

\section{Summary}
\label{sec_concl}

Usually the Yukawa matrix of the charged leptons is assumed to be diagonal.
However, it is known that this does not necessarily have to be the case.
In order to see how this assumptions affect neutrino observables 
we studied the impact of 
a non-diagonal charged lepton Yukawa matrix
on the neutrino sector of split 
supersymmetric models. This was done by using two different ans\"atze
for the charged lepton matrix.
It was found that the mass differences between the different neutrino 
species are effectively insensitive to the charged lepton sector. This 
confirms the usual assumption of a diagonal charged lepton matrix with 
this respect. However, when studying the neutrino mixing angles it was 
found that the form of the mass matrix of the charged leptons indeed can
provoke significant changes in the observables. We found that especially 
the solar and reactor mixing angles are sensible to this generalization,
whereas the atmospheric angle shows a somewhat weaker dependence.
Thus, it has been shown that the usual assumption of a diagonal mass matrix 
for charged leptons, can lead to important mistakes in the interpretation of 
experimental data. In other words, within a given model a parameter point 
that agrees with the experimental neutrino data in the context of a diagonal 
charged lepton matrix, is likely to disagree with the data in the context of 
a non-diagonal charged lepton matrix or viceversa.


\begin{acknowledgments}
{\small 
The work of M.A.D. was partly funded by Conicyt grant 1100837 (Fondecyt Regular). 
B.K. was funded by Conicyt-PBCT grant PSD73.}
\end{acknowledgments}


\end{document}